\documentclass[12pt]{article}
\usepackage{graphics,cite,amssymb,epsfig,float,psfrag}
\RequirePackage{epsfig}
\RequirePackage{subfigure}
\newcommand{\beq}{\begin{equation}}
\newcommand{\eequ}{\end{equation}}
\newcommand{\eeq}{\end{equation}}
\def\bea{\begin{eqnarray}}
\def\eea{\end{eqnarray}}
\def\as{\relax\ifmmode\alpha_s\else{$\alpha_s${ }}\fi}
\def \pt{\relax\ifmmode{p_t}\else{$p_t${ }}\fi}
\def\nn{\nonumber}
\newcommand{\noi}{\noindent}

\usepackage{epsfig}
\def\c{C}
\def\cs{{\c_7}}
\def\cn{{\c_9}}
\def\ct{{\c_{10}}}
\def\cne{\cn^{\rm eff}}
\def\cse{\cs^{\rm eff}}

\def\gl{\Gamma}

\def\l{\ell}

\def\d{{\rm d}}

\def\mh{\hat{m}}
\def\mbh{\mh_b}

\def\mvh{\mh_{K^*}}

\def\sh{\hat{s}}

\def\a{{\cal A}}

\def\ap{{A^\prime}}

\def\uh{{\hat{u}}}

\def\as{{\alpha_s}}
\def\be{\begin{equation}}
\def\ee{\end{equation}}
\def\ba{\begin{eqnarray}}
\def\ea{\end{eqnarray}}

\def\az    {\ensuremath{{\cal A}_{0}(s)}}
\def\ap    {\ensuremath{{\cal A}_{\parallel}(s)}}
\def\at    {\ensuremath{{\cal A}_{\perp}(s)}}

\def\azd   {\ensuremath{|\az|^{2}}}  
\def\apd   {\ensuremath{|\ap|^{2}}}
\def\atd   {\ensuremath{|\at|^{2}}}

\def\thetakstar {\ensuremath{\theta_{K^*}}}
\def\phitr      {\ensuremath{\phi_{tr}}}   
\def\thetatr    {\ensuremath{\theta_{tr}}}
\def\cthetatr   {\ensuremath{\cos{\thetatr}}}

\def\cthetakstar{\ensuremath{\cos{\thetakstar}}}

\def\sphitr     {\ensuremath{\sin{\phitr}}}
\def\pipt       {\ensuremath{\text{Im}{({\cal A}_{\parallel}^*(s)\at)}}}
\def\przp       {\ensuremath{\text{Re}{({\cal A}_{0}^*(s)\ap)}}}
\def\pizt       {\ensuremath{\text{Im}{({\cal A}_{0}^*(s)\at)}}}

\def\jpsi     {\ensuremath{ J/\psi }}
\def\dd          {\ensuremath{d }}   

\newcommand{\cq}[1]{\cos^{2}{#1}}
\newcommand{\sq}[1]{\sin^{2}{#1}}
\newcommand{\gfrac}[2]{\displaystyle\frac{#1}{#2}}
\newcommand{\text}[1]{\mbox{\rm #1}}

\newif\ifdtup

\def\eqal2#1{\,\vcenter{\openup1\jot
\caja   \ialign{\strut \hfil$\displaystyle{##}$&\hfil$
\displaystyle{{}##}$\hfil &$
\displaystyle{{}##}$\hfil\crcr#1\crcr}}\,}

\begin{document}

\title{
\begin{flushright}\normalsize
\vspace{0cm}
DESY 02-005 \\ 
hep-ph/0205254\\
May 2002
\vspace{0.cm}
\end{flushright}
\begin{center}
\centerline{\large\bf Helicity Analysis of the Decays $B \to K^* \ell^+ 
\ell^-$ and}
\vspace*{2mm}
\centerline{\large\bf $B \to \rho \ell \nu_\ell$ in the Large Energy 
Effective Theory}
\end{center}
}
\vspace{0.cm}

\author{Ahmed Ali\footnote{e-mail~:ahmed.ali@desy.de} ~and A. Salim 
Safir\footnote{e-mail~: safir@mail.desy.de} \\
        Deutsches Elektronen-Synchrotron DESY, \\
        D-22603 Hamburg, Germany.}
\par \maketitle

\begin{abstract}
We calculate the independent helicity amplitudes in the decays $B \to K^*
\ell^+ \ell^-$ and $B \to \rho \ell \nu_\ell$ in the so-called
Large-Energy-Effective-Theory (LEET). Taking into account the dominant
$O(\alpha_s)$ and $SU(3)$ symmetry-breaking effects, we calculate 
various Dalitz distributions in these decays making use of the presently
available data and decay form factors calculated in the QCD sum
rule approach. Differential decay rates in the dilepton 
invariant mass and the Forward-Backward asymmetry in $B \to K^* \ell^+ \ell^-$
are worked out. We also present the decay amplitudes 
in the transversity basis which has been used in the analysis of data on
the resonant decay $B \to K^* J/\psi (\to \ell^+ \ell^-)$. 
Measurements of the ratios $R_i(s) \equiv d \Gamma_{H_i}(s)(B \to K^*
\ell^+ \ell^-)/ d \Gamma_{H_i}(s)(B \to \rho \ell \nu_\ell)$, involving
the helicity amplitudes $H_i(s)$, $i=0,+1, -1$, as precision tests of
the standard model in semileptonic rare $B$-decays are emphasized. We
argue that $R_0(s)$ and $R_{-}(s)$ can be used to determine the CKM ratio
$\vert V_{ub}\vert/\vert V_{ts} \vert$ and search for new physics, where
the later is illustrated by supersymmetry.
\end{abstract}

\newpage \pagestyle{plain}

\section{Introduction}
\vskip 5mm 

Rare $B$ decays involving flavour-changing-neutral-current (FCNC) 
transitions, such as $b \to s \gamma$ and $b \to s \ell^+ \ell^-$, have
received a lot of theoretical interest \cite{Greub:1999sv}. First 
measurements of the decay $B \to X_s \gamma$ were reported by the
CLEO collaboration \cite{Alam:1995aw}. These decays
are now being investigated more precisely in experiments at the B 
factories. The current world average based on the improved measurements by
the CLEO \cite{Chen:2001fj}, ALEPH \cite{alephbsg}
and BELLE \cite{bellebsg} collaborations, 
${\cal B}(B \to X_s \gamma)=(3.22 \pm 0.40) \times 10^{-4}$, 
is in good agreement with the estimates of the standard model (SM) 
\cite{Chetyrkin:1997vx,Kagan:1999ym,Gambino:2001ew}, which we shall take
as ${\cal B}(B \to X_s \gamma)=(3.50 \pm 0.50) \times 10^{-4}$, reflecting
the parametric uncertainties dominated by the scheme-dependence of the
quark masses. The decay $B \to X_s \gamma$ also provides useful 
constraints on the parameters of the supersymmetric theories, which in the
context of the minimal supersymmetric standard model (MSSM) have been
recently updated in \cite{Ali:2001jg}.

 Exclusive decays involving the $b \to s \gamma$ transition
are best exemplified by the decay $B \to K^* \gamma$, which have been 
measured with a typical accuracy of $\pm 10\%$, the current branching 
ratios being \cite{Chen:2001fj,TajimaH:2001,Aubert:2001}
${\cal B}(B^\pm \to K^{*\pm} \gamma)=(3.82 \pm 0.47) \times 10^{-5}$   
and ${\cal B}(B^0 \to K^{* 0} \gamma)=(4.44 \pm 0.35) \times 10^{-5}$.
These decays have been analyzed recently
\cite{Ali:2001ez,Beneke:2001at,Bosch:2001gv}, by taking into account 
$O(\alpha_s)$ corrections, henceforth referred to as the
next-to-leading-order (NLO) estimates,  in
the large-energy-effective-theory (LEET) limit  
\cite{Dugan:1990de,Charles:1998dr}. As this framework does not predict the
decay form  factors, which have to be supplied from outside, 
consistency of NLO-LEET estimates with current data constrains
the magnetic moment form factor in $B \to K^* \gamma$  in 
the range $T_1^{K^*}(0)=0.27 \pm 0.04$. These values are somewhat lower than 
the corresponding estimates in the lattice-QCD framework, yielding  
\cite{DelDebbio:1997kr} $T_1^{K^*}(0)=0.32^{+0.04}_{-0.02}$, and
in the light cone QCD sum rule approach, which give typically
$T_1^{K^*}(0)=0.38 \pm 0.05$ \cite{Ball:1998kk,Ali:1999mm}. (Earlier
lattice-QCD results on $B \to K^* \gamma$ form factors are reviewed in
\cite{Soni:1995qq}.) It is  imperative to check the consistency of the
NLO-LEET estimates, as this
would provide a crucial test of the ideas on QCD-factorization, which have 
been formulated in the context of non-leptonic exclusive $B$-decays 
\cite{Beneke:1999br}, but which have also been invoked in the study of
exclusive radiative and semileptonic $B$-decays 
\cite{Ali:2001ez,Beneke:2001at,Bosch:2001gv}. The decays $B \to \rho 
\gamma$ and $B \to K^* \gamma$ provide a 
good consistency check of this framework, with the 
branching ratios, the isospin-violating ratio $\frac{\Gamma (B^\pm \to 
\rho^\pm \gamma)}{2 \Gamma(B^0 \to \rho^0 \gamma)}-1$ and direct 
CP-violating asymmetries, such as ${\cal A}(\rho^\pm \gamma) \equiv
\frac{{\cal B}(B^- \to \rho^- \gamma) -{\cal B}(B^+ \to \rho^+ \gamma)}
{{\cal B}(B^- \to \rho^- \gamma) + {\cal B}(B^+ \to \rho^+ \gamma)}$, being 
the quantities of interest \cite{Bosch:2001gv,Ali:2001ez}. Likewise, 
isospin-violation in the decays $B \to K^* \gamma$, defined as
$\Delta_{0-}= \frac{\Gamma(B^0 \to K^{*0} \gamma) -\Gamma(B^- \to K^{*-} 
\gamma)}{\Gamma(B^0 \to K^{*0} \gamma) + \Gamma(B^- \to K^{*-}\gamma)}$
and its charge conjugate $\Delta_{0+}$, will also test this framework
\cite{Kagan:2001zk}.

The exclusive decays $B \to K^* \ell^+ \ell^-$, $\ell^\pm =e^\pm,\mu^\pm$    
have also been studied in the NLO-LEET approach in
\cite{Beneke:2001wa,Beneke:2001at}.
 In this case, the LEET symmetry brings an enormous
simplicity, reducing the number of independent form factors from seven to
only two, corresponding to the transverse and longitudinal polarization of 
the virtual photon in the underlying process $B \to K^* \gamma^*$,
called hereafter $\xi_\perp^{K^*}(q^2)$ and $\xi_{||}^{K^*}(q^2)$. The
same symmetry reduces the number of independent form factors in the decays
$B \to \rho \ell \nu_\ell$ from four to two. Moreover, in the $q^2$-range
where the large energy limit holds, the two set of form factors are equal 
to each other, up to $SU(3)$-breaking corrections, which are already 
calculated in specific theoretical frameworks. Thus, knowing $V_{ub}$
precisely, one can make theoretically robust predictions for the 
rare $B$-decay $B \to K^* \ell^+ \ell^-$ from the measured $B \to \rho 
\ell \nu_\ell$ decay in the SM. The LEET symmetries are broken by QCD 
interactions and the leading $O(\alpha_s)$ corrections in perturbation 
theory are known \cite{Beneke:2001wa,Beneke:2001at}. We make use of 
these theoretical developments and go a step further in that we calculate
the various independent helicity amplitudes in the decays $B \to K^* 
\ell^+ \ell^-$ and $B \to \rho \ell \nu_\ell$ in the NLO accuracy
in the large energy limit. We recall that a  
decomposition of the final state $B \to K^* (\to K \pi) \ell^+ \ell^-$
in terms of the helicity amplitudes $H_\pm^{L,R}(q^2)$ and 
$H_0^{L,R}(q^2)$, without the explicit $O(\alpha_s)$ corrections, was
undertaken in a number of papers
\cite{Melikhov:1998cd,Aliev:1999gp,Kim:2000dq,Kim:2001xu,Nguyen:2001zu,Chen:2002bq}.
In particular, Kim et al.~\cite{Kim:2000dq,Kim:2001xu}  
emphasized the role of the azimuthal angle distribution as a
precision test of the SM. Following closely the earlier analyses, we now
calculate the $O(\alpha_s)$ corrections in the LEET framework.

Concentrating on the decay $B \to K^* \ell^+ \ell^-$, the main theoretical
tool is the factorization Ansatz which enables one to relate the form 
factors in full QCD (called in the literature $A_0(q^2), A_1(q^2), A_2(q^2), 
V(q^2), T_1(q^2), T_2(q^2), T_3(q^2)$) and the two LEET form factors  
 $\xi_\perp (q^2)$ and $\xi_{||}(q^2)$ 
\cite{Beneke:2001wa,Beneke:2001at};
\begin{equation}
f_k(q^2)= C_{\perp}\xi_\perp (q^2) + C_{||}\xi_{||}(q^2) + \Phi_B \otimes 
T_k \otimes \Phi_V~,
\label{eq:fact}
\end{equation}
where the quantities $C_i$ $(i=\perp, \parallel)$ encode the perturbative 
improvements of the factorized part
\begin{equation}
C_i=C_i^{(0)} + \frac{\alpha_s}{\pi} C_i^{(1)} + ... ,
\nonumber
\end{equation}
and $T_k$ is the hard spectator kernel (regulated so as to be free of 
the end-point singularities), representing the 
non-factorizable perturbative corrections, with the direct product
understood as a convolution of $T_k$ with the light-cone distribution 
amplitudes of the $B$ meson ($\Phi_B$) and the vector meson ($\Phi_V$). 
With this Ansatz, it is a straightforward exercise to implement the
$O(\alpha_s)$-improvements in the various helicity amplitudes. 
The non-perturbative information is encoded in the LEET-form factors,
which are {\it a priori} unknown, and the various parameters which enter 
in the description of the non-factorizing hard spectator contribution,
which we shall discuss at some length. The normalization of the LEET form
factor $\xi_\perp^{K^*} (q^2)$ at $q^2=0$ is determined by the
$B \to K^* \gamma$ decay rate; the other form factor $\xi_{||}^{K^*}(q^2)$
has to be modeled entirely for which we
use the light cone QCD sum rules. This input, which
for sure is model-dependent, is being used to illustrate the various
distributions and should be replaced as
more precise data on the decay $B \to \rho \ell \nu_\ell$ becomes
available, which then can be used directly to determine 
the form factors $\xi_\perp^{K^*} (q^2)$ and $\xi_{||}^{K^*} (q^2)$,
taking into account the $SU(3)$-breaking effects.

 Using the effective Hamiltonian approach, and incorporating the 
perturbative improvements, we calculate a number of Dalitz distributions,
the dilepton invariant mass distribution for the individual helicity 
amplitudes (and the sum), and the forward-backward asymmetry in
$B \to K^* \ell^+ \ell^-$. As the range of validity of the LEET-based
estimates in this decay is restricted to the large-$E_{K^*}$ region, we
shall restrict ourselves to the low $s$-region in the dilepton invariant
mass, which for the sake of definiteness is taken as $s \leq 8$ GeV$^2$.
We shall also neglect the contributions from the long-distance effects
to the final state $B \to K^* \ell^+ \ell^-$, arising from the process
$B \to K^* (\rho, \omega, \phi) \to K^* \ell^+ \ell^-$, as they are
expected to be tiny due to the CKM-suppression and the small
leptonic branching ratios of the vector mesons $\rho, \omega, \phi$.
To project out the various helicity components
experimentally, one can use the Dalitz distribution in the 
dilepton invariant mass ($s=q^2$) and $\cos \theta_K$, where $\theta_K$ is 
the polar angle of the $K$ meson in the rest system of the $K^*$ meson
measured with respect to the helicity axis, i.e., the outgoing direction 
of the $K^*$. The angular distribution allows to separate the $0$-helicity 
component $\vert H_0(s) \vert^2 = \vert H_0^L(s) \vert^2 + \vert H_0^R(s) 
\vert^2$ and the sum $\vert H_+(s) \vert^2 + \vert H_-(s) \vert^2$. In the
SM, and other beyond-the-SM scenarios considered here which have the 
same operator basis, the component $H_+(s)$ is negligibly small.
This holds for both the left-handed and right-handed projections,
$H_+^L(s)$ and $H_+^R(s)$. We show this here in the case of the SM. Hence,
for all practical considerations, these components can be ignored 
and we concentrate on the $H_-(s)$ and $H_0(s)$ components.
We show the systematic improvements in $O(\alpha_s)$ and
$1/M$ in $H_{-}(s)$ and $H_{0}(s)$ in these decays. Their measurements,     
in conjunction with the decay distributions in $B \to \rho \ell \nu_\ell$,   
will serve as precision tests of the flavour sector in the
SM, yielding $\vert V_{ub}\vert/\vert V_{ts}\vert$, and in searching for
possible deviations from the SM, exemplified here by supersymmetry.        

We also work out the decay amplitudes for $B \to K^*(\to K \pi)
\ell^+ \ell^-$ in the transversity basis
 \cite{ref:dunietz,Kramer:1991xw,ref:dighe},
which has been used by several experimental groups to measure the
corresponding amplitudes for the decay $B \to K^* (\to K \pi) J/\psi (\to
\ell^+ \ell^-)$ \cite{ref:cleo,ref:cdf,ref:babar,ref:belle}. These involve
the complex amplitudes ${\cal A}_0(s)$, ${\cal A}_\parallel (s)$ and
${\cal A}_\perp(s)$. The
amplitudes in the transversity and helicity bases are simply related
\cite{Kruger:1999xa} and,
having worked them out in the helicity basis, it is a straightforward
numerical exercise to work out the moduli and arguments of the amplitudes
in the transversity basis. Restricting ourselves to low-$s$ region ($s
\leq 8$ GeV$^2$), we show the results using the LEET approach both in the
LO and NLO. For illustrative purpose, we show the amplitudes in the
entire kinematically allowed region in the LO. The LEET-based transversity
amplitudes for the decay $B \to K^* \ell^+ \ell^-$ are
found to be in reasonable agreement with their measured counterparts in
the resonant decay $B \to K^* J/\psi (\to \ell^+ \ell^-)$.
Measurement of the short-distance component of these
amplitudes coming from  $B \to K^* \ell^+ \ell^-$ away from
$s=m_{J/\psi}^2$, in particular in the region $4 m_\ell^2 \leq s <
m_{J/\psi}^2$, will test the underlying LEET-based framework.

This paper is organized as follows: In section 2, we define the 
effective Hamiltonian and the matrix element for the decay
$b \to s \ell^+ \ell^-$. In section 3, we discuss the form factors in 
the LEET approach for the decay $B \to K^* \ell^+ \ell^-$, borrowing
heavily from the literature  
\cite{Charles:1998dr,Beneke:2001wa,Beneke:2001at},
give parametrizations for the two remaining form factors 
$\xi^{K^*}_{\perp}(s)$ and $\xi^{K^*}_{\parallel}(s)$
and specify other input parameters in our analysis. In section 4, we 
introduce the helicity amplitudes $H_{\pm}^{L,R}(s)$ and $H_{0}^{L,R}(s)$, 
give the $O(\alpha_s)$-improved expressions for these amplitudes and 
write down the Dalitz distributions in the set of variables $(\phi,s)$,
$(\cos \theta_K, s)$, and $(\cos \theta_{+},s)$. The quantities
$\vert H_{\pm}^{L,R}(s)\vert^2$ are shown as functions of $s$. Likewise, 
Dalitz distributions in $(\cos \theta_{+},s)$ are shown for the two
dominant components, $H_{0}(s)$ and $H_{-}(s)$, and adding all three
components. We also show the dilepton invariant mass distributions
for the individual helicity amplitudes, and their sum, and the forward 
backward asymmetry, making explicit the $O(\alpha_s)$ improvements. 
Section 5 describes the amplitude decomposition for $B \to K^* \ell^+
\ell^-$ in the transversity basis. We show the amplitudes
$\vert {\cal A}_{0}(s) \vert^2$, $\vert {\cal A}_{\parallel}(s) \vert^2$
and
$\vert {\cal A}_{\perp}(s) \vert^2$, as well as the relative phases
$\phi_{\parallel}(s)$ and $\phi_{\perp}(s)$, making explicit the
$O(\alpha_s)$ improvements in these quantities. Extrapolating the LO
results for these quantities to the $J/\psi$ mass,  
we compare them with data on $B \to K^* J/\psi (\to \ell^+ 
\ell^-)$.
In section 6, we turn to the decay distributions in the decay $B
\to \rho \ell \nu_\ell$, and display the various helicity components,
Dalitz distributions, and the dilepton ($\nu_\ell \ell$) invariant mass.
Estimates of the $B \to \rho$ LEET form factors
$\xi^{\rho}_{\perp}(s)$ and $\xi^{\rho}_{\parallel}(s)$, which are
scaled from their $B \to K^*$ counterparts incorporating SU(3)-breaking,
are also displayed here.
Section 7 is devoted to the determination of the ratio of the CKM matrix
elements $\vert V_{ub}\vert /\vert V_{ts}\vert$
from the ratio of the dilepton mass spectra in $B \to \rho \ell 
\nu_\ell$ and $B \to K^* \ell^+ \ell^-$ decays involving definite 
helicity states. In particular, we show the dependence of the ratio 
$R_{-}(s) ={{d\Gamma_{H_-}^{B \rightarrow  K^{*} \ l^{+}  l^{-}}/ds}
\over{d\Gamma_{H_-}^{B \rightarrow  \rho \ l \bar{\nu}}/ds}}$
and $R_0(s)$, involving the helicity-0 components, on the
CKM matrix elements $\vert V_{ub} \vert/\vert V_{ts} \vert$. Section 8 is
devoted to an  analysis of the ratios $R_0(s)$ and $R_-(s)$ to probe for
new physics in the decay $B \to K^* \ell^+ \ell^-$,
and illustrate this using some specific supersymmetric scenarios. 
Finally, section 9 contains a summary and some concluding remarks.

\section{\bf Effective Hamiltonian for $b \to s \ell^+ \ell^-$}
\hspace*{\parindent}
At the quark level, the rare semileptonic decay 
$b \to s  \ell^+ \ell^-$ can be described in terms of the effective 
Hamiltonian obtained by integrating out the top quark and $W^\pm$ 
bosons: 
\begin{equation}
H_{eff} = - \frac{G_F}{\sqrt{2}}  V_{t s}^\ast  V_{tb}  
\sum_{i=1}^{10}  C_{i}(\mu)  {\cal O}_i(\mu) \; , 
        \label{eq:he}
\end{equation}
\noi where $V_{ij}$ are the CKM matrix elements 
\cite{ckm} and  $G_F$ is the Fermi coupling constant. We use the 
operator  basis introduced in \cite{Chetyrkin:1997vx} for the operators 
${\cal O}_i$, $i=1,\ldots,6$, and define:

\begin{equation}
{\cal O}_7=-\frac{g_{em} \ m_b}{8\pi^2}\,\bar{s}\ \sigma^{\mu\nu}
(1+\gamma_5)\ b\ F_{\mu\nu},\qquad
{\cal O}_8=-\frac{g_s m_b}{8\pi^2}\,\bar{s}_i\sigma^{\mu\nu}
(1+\gamma_5)T^a_{ij}b_j G^a_{\mu\nu},
\end{equation}
\begin{equation}
{\cal O}_{9,10} = \frac{\alpha_{em}}{2\pi}\,
(\bar{\ell} \ell)_{V,A}\,(\bar{s}\ b)_{V-A},
\end{equation}
\noi where $\alpha_{em}=g^2_{em}/4 \pi$ is the
electromagnetic fine-structure constant.  $T^{a}$, $a=1....8$ are the
generators of QCD, and $i,\
j$ are $SU(3)$ color indices. Here $F_{\mu \nu}$ and $G^{a}_{\mu \nu}$
denote the electromagnetic and chromomagnetic field strength tensor,
respectively. 
The above Hamiltonian leads to the following free quark decay amplitude: 
\begin{eqnarray}
M(b\to s\ell^+\ell^-) & = & \frac{G_F \alpha_{em}}{\sqrt{2}  \pi} \ V_{t s}^{*} V_{tb} \, 
\Big\{ C_{9}  \left[ \bar{s}  \gamma_{\mu}  L  b \right] \left[ \bar{l}  \gamma^{\mu}  \l \right]
+ C_{10} \left[ \bar{s} \gamma_{\mu}  L  b \right] \left[ \bar{l}
\gamma^\mu  \gamma_5  \l \right] 
\nn\\
& & \; \; \; \; \; \; \; \; \; \; \; \; \; \; 
\; \; \; \; \; \; \; \; \; \left. 
- 2 \hat{m_{b}}  C_{7}^{\bf eff}  \left[ \bar{s}  i  \sigma_{\mu \nu}  {\hat{q^{\nu}}\over\hat{s}} R  b \right] 
\left[ \bar{l}  \gamma^{\mu}  \l \right] \right \} \; .
\label{eq:m}
\end{eqnarray}
Here, $L/R \equiv {(1 \mp \gamma_5)}/2$, $s=q^2$, $\sigma_{\mu \nu}={i\over
2}[\gamma_{\mu},\gamma_{\nu}]$ and $q_{\mu}=(p_{+} +p_{-})_{\mu}$,
where $p_{\pm}$ are the four-momenta of the leptons.
We put $m_s/m_b = 0$ and the hat denotes normalization
in terms of the $B$-meson mass, $m_B$, e.g. $\hat{s}=s/m_B^2$,
$\hat{m}_b =m_b/m_B$. Here  and in the remainder of this work we shall
denote by $m_b \equiv m_b(\mu)$ the $\overline{MS}$ mass evaluated at a
scale $\mu$, and by $m_{b,pole}$ the pole mass of the $b$-quark. To
next-to-leading order the pole and $\overline{\rm MS}$ masses are related by 
\begin{equation}
\label{polerel}
m_b(\mu) = m_{b,pole}\left(1+\frac{\alpha_s(\mu) C_F}{4\pi}\left[
3\ln\frac{m_b^2}{\mu^2}-4\right]+O(\alpha_s^2)\right)~.
\end{equation}
Since we are including the next-to-leading corrections into our
analysis, we will take the Wilson coefficients in
next-to-leading-logarithmic order (NLL) given in Table \ref{tab1}.
\begin{table}[t]
\vspace{0.1cm}
\begin{center}
\begin{tabular}{|l|c|c|c|c|c|c|}
\hline\hline
\rule[-2mm]{0mm}{7mm}
 & $\bar{C_1}$ & $\bar{C_2}$ & $\bar{C_3}$ & $\bar{C_4}$ & $\bar{C_5}$ & $\bar{C_6}$ \\
\hline
\rule[-0mm]{0mm}{4mm}
LL    & $-0.257$ & $1.112$ & $0.012$ & $-0.026$ & $0.008$ & $-0.033$ \\

NLL   & $-0.151$ & $1.059$ & $0.012$ & $-0.034$ & $0.010$ & $-0.040$ \\
\hline
\rule[-2mm]{0mm}{7mm}
 & $C_7^{\rm eff}$ & $C_8^{\rm eff}$ & $C_9$ & $C_{10}$
 & $C_9^{\rm NNLL}$ &  $C_{10}^{\rm NNLL}$ \\
\hline
\rule[-0mm]{0mm}{4mm}
LL  & $-0.314$ & $-0.149$ & $2.007$ & 0
 & & \\
NLL & $-0.308$ & $-0.169$ & $4.154$ & $-4.261$
 & \raisebox{2.5mm}[-2.5mm]{$4.214$} & \raisebox{2.5mm}[-2.5mm]{$-4.312$} \\
\hline\hline
\end{tabular}
\end{center}
\centerline{\parbox{14cm}{\caption{\label{tab1}
 Wilson coefficients at the scale $\mu=4.6\,$GeV in
leading-logarithmic (LL) and next-to-leading-logarithmic order
(NLL) \cite{Beneke:2001at}.}}}
\end{table}
%
\section{Form factors in the Large Energy Effective Theory}
\hspace*{\parindent}
Exclusive decays $B\to K^* \ell^+ \ell^-$ are
described by the matrix elements of the quark operators in
Eq.~(\ref{eq:m}) over meson states, which can be parameterized in terms of form
factors.

For the vector meson $K^*$ with polarization vector $\epsilon_\mu$,
the semileptonic form factors of the $V-A$ current are defined as
\begin{eqnarray}
&&\langle{K^* (p,\epsilon^*)} | (V-A)_\mu | B(p_B)\rangle  = -i\
\epsilon^*_\mu 
(m_B+m_{K^*}) A_1(s) + i\ (p_B+p)_\mu (\epsilon^* p_B)\,
\frac{A_2(s)}{m_B+m_{K^*}}\nn\\
&& +  i\ q_\mu (\epsilon^* p_B) \,\frac{2m_{K^*}}{s}\,
\Big(A_3(s)-A_0(s) \Big) +\epsilon_{\mu\nu\rho\sigma}\ \epsilon^{*\nu} p_B^\rho p^\sigma\,
\frac{2V(s)}{m_B+m_{K^*}} ~.
\hspace*{2cm}\label{eq:ff3}
\end{eqnarray}
Note the exact relations:
\begin{eqnarray}
 A_3(s) & = & \frac{m_B+m_{K^*}}{2m_{K^*}}\, A_1(s) -
\frac{m_B-m_{K^*}}{2m_{K^*}}\, A_2(s),\nonumber\\
A_0(0) & = & A_3(0), \nonumber\\
\langle K^* |\partial_\mu A^\mu | B\rangle & = & 2 m_{K^*}
(\epsilon^* p_B) A_0(s).
 \label{eq:A30}
\end{eqnarray}
The second relation in (\ref{eq:A30}) ensures that there is no kinematical
singularity in the matrix element at $s=0$. The decay $B\to
K^*\ell^+\ell^-$ is described by the above semileptonic form factors and the
 following penguin form factors:
\begin{eqnarray}
\langle { K^* (p,\epsilon^*)} | C_7^{\,\rm eff}
\bar{s}\sigma_{\mu\nu}q^{\nu}(1+\gamma_5)b |B(p_B)\rangle & = & i\ \epsilon_{\mu\nu\rho\sigma} \epsilon^{*\nu}
p_B^\rho p^\sigma \, 2 {\cal T}_1(s)\label{eq:T}\\
&+&  {\cal T}_2(s)  \Big\{ \epsilon^*_\mu (m_B^2-m_{{K^*}}^2) 
- (\epsilon^* p_B) \,(p_B+p)_\mu  \Big\}\nn\\
&+ &  {\cal T}_3(s)(\epsilon^* p_B) 
\left \{ q_\mu - \frac{s}{m_B^2-m_{{K^*}}^2}\ (p_B+p)_\mu \right \} ~.\nn
\end{eqnarray}
The matrix element decomposition is defined such that the leading
order contribution from the electromagnetic dipole operator ${\cal
O}_7$ reads ${\cal T}_i(s) = C^{\bf eff}_7 T_i(s)+\ldots$, where $T_i(s)$ denote the tensor form factors. Including also the four-quark operators (but neglecting for the moment annihilation contributions), the leading logarithmic expressions are \cite{Grinstein:1989me}

\begin{eqnarray}
\label{calT1}
{\cal T}_1(s) &=& C_7^{\,\rm eff} \,T_1(s) + Y(s) \,\frac{s}
{2 m_b (m_B+m_{K^*})}\,V(s), \\
\label{calT2}
{\cal T}_2(s) &=& C_7^{\,\rm eff} \,T_2(s) + Y(s) \,\frac{s}
{2 m_b (m_B-m_{K^*})}\,A_1(s), \\
{\cal T}_3(s) &=& C_7^{\,\rm eff} \,T_3(s) + Y(s) \,\left[\frac{m_B-m_{K^*}}{2 m_b} \,A_2(s)- \frac{m_B+m_{K^*}}{2 m_b}\,A_1(s)\right], 
\label{calT3}
\end{eqnarray}

\noi with $C_7^{\,\rm eff} = C_7-C_3/3-4 C_4/9-20 C_5/3-80 C_6/9
=C_7-(4 \bar{C}_3-\bar{C}_5)/9-(4 \bar{C}_4-\bar{C}_6)/3$, and 
\begin{eqnarray}
\label{yy}
Y(s) &=& h(s,m_c) \left(3 \bar{C}_1+\bar{C}_2+3 \bar{C}_3+\bar{C}_4+3
 \bar{C}_5+\bar{C}_6\right) \nonumber\\
&&-\,\frac{1}{2}\,h(s,m_b) \left(4 \,(\bar{C}_3+\bar{C}_4)+3 \bar{C}_5+
\bar{C}_6\right) -\frac{1}{2}\,h(s,0) \left(\bar{C}_3+3 \bar{C}_4\right)
\nonumber\\
&&+\,\frac{2}{9}\,\left(\frac{2}{3}\bar{C}_3+2 \bar{C}_4+\frac{16}{3} 
 \bar{C}_5\right)~,
\end{eqnarray}
\noi where the ``barred'' coefficients $\bar{C}_i$ ( for i=1,...,6) are
defined as certain linear combinations of the $C_i$, such that the
$\bar{C}_i$ coincide  {\it at leading logarithmic order} with the
Wilson coefficients in the standard basis \cite{Buchalla:1996vs}.
Following Ref. \cite{Beneke:2001at}, they are expressed as~:
 \begin{eqnarray}
\bar{C}_1 &=& \frac{1}{2} \,C_1,\nonumber\\
\bar{C}_2 &=& C_2-\frac{1}{6}\,C_1,\nonumber\\
\bar{C}_3 &=& C_3-\frac{1}{6}\,C_4+16\,C_5-\frac{8}{3}\,C_6,\nonumber\\
\bar{C}_4 &=& \frac{1}{2}\,C_4+8\,C_6,\nonumber\\
\bar{C}_5 &=& C_3-\frac{1}{6}\,C_4+4\,C_5-\frac{2}{3}\,C_6,\nonumber\\
\bar{C}_6 &=& \frac{1}{2}\,C_4+2\,C_6.
\end{eqnarray}

\noi The function
\begin{equation}
h(s,m_q) = -\frac{4}{9}\left(\ln\frac{m_q^2}{\mu^2} - \frac{2}{3} - z \right)-
\frac{4}{9} \,(2+z) \,\sqrt{\,|z-1|} \,
\left\{\begin{array}{l}\,\arctan\displaystyle{\frac{1}{\sqrt{z-1}}}\qquad\quad 
z>1~,\\
[0.4cm]\,\ln\displaystyle{\frac{1+\sqrt{1-z}}{\sqrt{z}}} - \frac{i\pi}{2}
\quad z\leq 1~,\end{array}\right.
\end{equation}
\noi is related to the basic fermion loop. (Here $z$ is defined as
$4 m_q^2/s$.) $Y(s)$ is given in the NDR scheme with anticommuting
$\gamma_5$ and with respect to the operator basis of
\cite{Chetyrkin:1997vx}.
Since $C_9$ is basis-dependent starting from next-to-leading logarithmic
order, the terms not proportional to $h(s,m_q)$ differ from those given in
\cite{Buchalla:1996vs}.
The contributions from the four-quark operators ${\cal O}_{1-6}$ are
usually combined with the coefficient $C_9$ into an ``effective''
(basis- and scheme-independent) Wilson coefficient
$C_9^{\,\rm eff}(s)=C_9+Y(s)$. 

Recently, it has been shown that the symmetries emerging in the large
energy limit \cite{Charles:1998dr} relate the otherwise independent
form factors entering in the decays of $B$ mesons
into light mesons. However this symmetry is restricted to the
kinematic region in which the energy of the final state meson scales
with the heavy quark mass. For the $B\to K^* \ell^+ \ell^-$ decay, this
region is identified as $s \simeq 8\ GeV^2$.

\noi Thus,  in the large energy limit, the standard form factors $ A_{0},
A_{1},A_{2}, V, T_{1}, T_{2}$ and $T_{3}$ can be expressed in terms of
two universal functions $\xi_{||}(s)$ and  $\xi_{\perp}(s)$
 \cite{Charles:1998dr}~:

\begin{eqnarray}
A_{0}(s) &=& (1 - {m_{V}^2\over m_{B} E_V })\ \xi_{||}(s) +
 {m_{V}\over m_B}\ \xi_{\perp}(s)~,\label{A0} \\
A_{1}(s) &=&{2 E_V \over m_{B}+m_{V}}\ \xi_{\perp}(s)~,\label{A1} \\
A_{2}(s)&=& (1 + {m_{V}\over m_B})\ [\ \xi_{\perp}(s) - {m_{V}\over E_V}\
 \xi_{||}(s)\ ]~,\label{A2} \\
V(s)&=&(1 + {m_{V}\over m_B})\ \xi_{\perp}(s)\label{V} \\
T_{1}(s)&=& \xi_{\perp}(s)~, \label{T1} \\
T_{2}(s)&=&(1- {s\over m_{B}^2-m_{V}^2}) \ \xi_{\perp}(s) ~,\label{T2} \\
T_{3}(s)&=& \xi_{\perp}(s)- {m_{V}\over E_V}\ (1- {m_{V}^2\over m_{B}^2}) \
 \xi_{||}(s)~,\label{T3} 
\end{eqnarray}
\noi where
\begin{eqnarray} 
E_V = {m_B\over 2} \Big (1 - {s\over m_B^2} + {m_V^2 \over m_B^2} \Big),  
\end{eqnarray}
refers to the energy of the 
final  vector meson V and $\xi_{\perp,\parallel}(s)$ refer to the form 
factors in the large energy limit (called subsequently as the LEET
form factors). However, these symmetries are broken by factorizable and
non-factorizable QCD corrections, worked out in the present context by Beneke 
{\it et al.} \cite{Beneke:2001at,Beneke:2001wa}. Since, we are using in
our analysis the definitions of the form factors 
$\xi_{\perp,||}(s)$ by Charles {\it et al.} \cite{ Charles:1998dr},
the factorizable corrections obtained in \cite{Beneke:2001wa} are
expressed as follows:
\begin{eqnarray}
\label{A1cor}
A_{1}(s)&=& {2 E_{V}\over m_{B}+m_{V}}\ \xi_{\perp}(s)+  {\alpha_{s}
 C_{F}\over 4 \pi} \ \Delta A_{1}~,\\
\label{A2cor}
A_{2}(s)&=& {m_{B} \over m_{B} - m_{V}} \left[ \xi_{\perp}(s) - {m_{V}\over
E_{V}} \ \xi_{||}(s) (1 +  {\alpha_{s} C_{F}\over 4 \pi} [-2 +2L])\right]+ 
{\alpha_{s} C_{F}\over 4 \pi} \ \Delta A_{2}~,
\end{eqnarray}
\noi with
\begin{eqnarray}
\label{L}
 L & =& - {2 E_V\over m_{B} -2E_V}\ \ln {2 E_V \over m_{B} }~,\nn\\ 
\Delta A_{1}=0,\,\,\Delta A_{2}&=& {m_{V} \over m_{B} - m_{V}}\ {m_{B}^2 ( m_{B}
 -2E_V) \over  4  E_V^3}  \Delta F_{||}~,\nn\\
\Delta F_{||}&=& {8 \pi^2 f_{B} f_{V}\over N_{C} m_{B}}
\langle \ell_{+}^{-1}\rangle_{+}
 \langle\bar{u}^{-1}\rangle_{||}\label{DeltaF||}~,
\end{eqnarray}
\noi where $f_{B}$ and $f_{V}$ are, respectively, the meson decay
constants for the B meson and the corresponding  V meson.  The above
expression for $\Delta A_2$ also involves a non-perturbative quantity
$\langle \ell_{+}^{-1}\rangle_{+}$. Formally, 
$\langle \ell_{+}^{-1}\rangle_{+}\sim {1\over \Lambda_{QCD}}$, but nothing
more is
known about this universal parameter at present.  It is estimated to lie
in the range $(0.2 - 0.5\ GeV)^{-1}$ \cite{Beneke:2001at}, following which we
take $(0.3\ GeV)^{-1}$ as our default value for this quantity in our
calculations. For the $K^{*}$ meson, we use the result quoted in
Ref.~\cite{Beneke:2001at} : $ <\bar{u}^{-1}>_{||}= 3.75$. Concerning the
form factors $A_0(s)$ and $V(s)$, defined respectively in Eqs.~(\ref{A0})
and (\ref{V}), they hold exactly to all orders in perturbations theory
and this defines the factorization scheme.

The remaining contributions arising from the hard 
spectator corrections for the $B \rightarrow \ V \ l^{+} \ l^{-}$ decay have
been computed recently by Beneke {\it et al.} \cite{Beneke:2001at}, yielding
\begin{eqnarray}
\label{tperpdef}
&& {\cal T}_1(s) \equiv {\cal T}_\perp(s) , \\
&&{\cal T}_2(s) = \frac{2 E_V}{m_B}\, {\cal T}_\perp(s), \\
&&{\cal T}_3(s) - \frac{m_B}{2 E_V}\,{\cal T}_2(s) \equiv  {\cal T}_\parallel(s) ,\label{tpardef}
\end{eqnarray}
with
\begin{eqnarray}
\label{nlodef}
{\cal T}_{\perp} &=& \xi_{\perp} \left(C_{\perp}^{(0)}+\frac{\alpha_s C_F}
{4\pi} \,C_{\perp}^{(1)}\right) \nonumber\\
&&\hspace*{0cm}+ \,\frac{\pi^2}{N_c}
\,\frac{f_B f_{V,\,{\perp}}}{m_B} \,\,\Xi_{\perp}\,\sum_{\pm}
\int\frac{d\omega}{\omega}\,\Phi_{B,\,\pm}(\omega)\int_0^1
\!du\,\Phi_{V,\,{\perp}}(u) \,T_{{\perp},\,\pm}(u,\omega),\nn\\
{\cal T}_{||} &=& \xi_{||} {m_V\over E_V} \left(C_{||}^{(0)}+\frac{\alpha_s C_F}
{4\pi} \,C_{||}^{(1)}\right) \nonumber\\
&&\hspace*{0cm}+ \,\frac{\pi^2}{N_c}
\,\frac{f_B f_{V,\,{||}}}{m_B} \,\,\Xi_{||}\,\sum_{\pm}
\int\frac{d\omega}{\omega}\,\Phi_{B,\,\pm}(\omega)\int_0^1
\!du\,\Phi_{V,\,{||}}(u) \,T_{{||},\,\pm}(u,\omega)~.
\end{eqnarray}
\noi Here $C_F=4/3$, $N_c=3$, $\Xi_\perp\equiv 1$, $\Xi_{||}
\equiv m_{V}/E_V$, and the hard-scattering term $T_{a,\,\pm}(u,\omega)$
 $(a=\perp,||)$ is expanded as~:
\begin{equation}
T_{a,\,\pm}(u,\omega)= T^{(0)}_{a,\,\pm}(u,\omega)+\frac{\alpha_s C_F}{4\pi} \,
T^{(1)}_{a,\,\pm}(u,\omega)~,
\end{equation}
\noi where $f_{K^*,\,||}$ denotes the usual $K^*$ decay constant $f_{K^*}$ and $f_{K^*,\,\perp}$ refers to the (scale-dependent) transverse decay constant defined by the matrix element of the tensor current. 

The coefficient $C_a^{(1)}$ $(a=\perp,||)$ in
(\ref{nlodef}) represents the next-to-leading order form factor
correction, and can be expressed as~:
\begin{eqnarray}
C_a^{(1)}=C_a^{(\bf f)}+C_a^{(\bf nf)}~,\label{Cafct}
\end{eqnarray}
\noi where $C_a^{(\bf f)}$ contains a factorizable term from expressing
the full QCD form factors in terms of $\xi_a$ in Eqs.~(\ref{calT1}),
(\ref{calT2}) and (\ref{calT3}). The non-factorizable correction
$C_a^{(\bf nf)}$ is obtained by computing matrix elements of
four-quark operators and the chromomagnetic dipole operator. 
The matrix elements of four-quark operators require the calculation
of two-loop diagrams, and the result for the current-current
operators ${\cal O}_{1,2}$ as well as the matrix element of the
chromomagnetic dipole operator can be extracted from
Ref.~\cite{AAGW}. The 2-loop matrix elements of the QCD penguin
operators have not yet been computed and hence will be
neglected. This should be a very good approximation due to the small
Wilson Coefficients of the penguin operators. For the definitions of 
the  parameters in Eq.~(\ref{nlodef}), we refer to \cite{Beneke:2001at}.

\begin{table}
\addtolength{\arraycolsep}{3pt}
\renewcommand{\arraystretch}{1.4}
$$
\begin{array}{|l|llll|lll|}
\hline
& A_1 & A_2 & A_0 & V & T_1 & T_2 & T_3\\ \hline
F(0) & 0.294 & 0.246 & 0.412 &0.399 & 0.334 & 0.334 & 0.234\\

c_1 & 0.656 & 1.237 & 1.543 & 1.537 & 1.575 & 0.562 & 1.230\\

c_2 & 0.456 & 0.822 & 0.954 & 1.123 & 1.140 & 0.481 & 1.089\\
\hline
\end{array}
$$
\caption[] {Input values for the parameterization 
  (\protect{\ref{eq:para}}) of the $B\to K^*$ form factors. Renormalization
   scale for
  the penguin form factors $T_i$ is $\mu = m_b$ 
\cite{Ali:1999mm}.} \label{tab:p1}
 \end{table}

\begin{figure}
\psfrag{a}{\hskip 0cm $s\ (GeV^2)$}
\psfrag{c}{\hskip -1.3cm $\xi^{(K^*)}_{||}(s)$}
\psfrag{b}{\hskip -1.3cm $\xi^{(K^*)}_{\perp}(s)$}
\psfrag{d}{\hskip 0cm [AS]}
\psfrag{e}{\hskip 0cm [BFS]}
\begin{center}
\includegraphics[width=18cm,height=16cm]{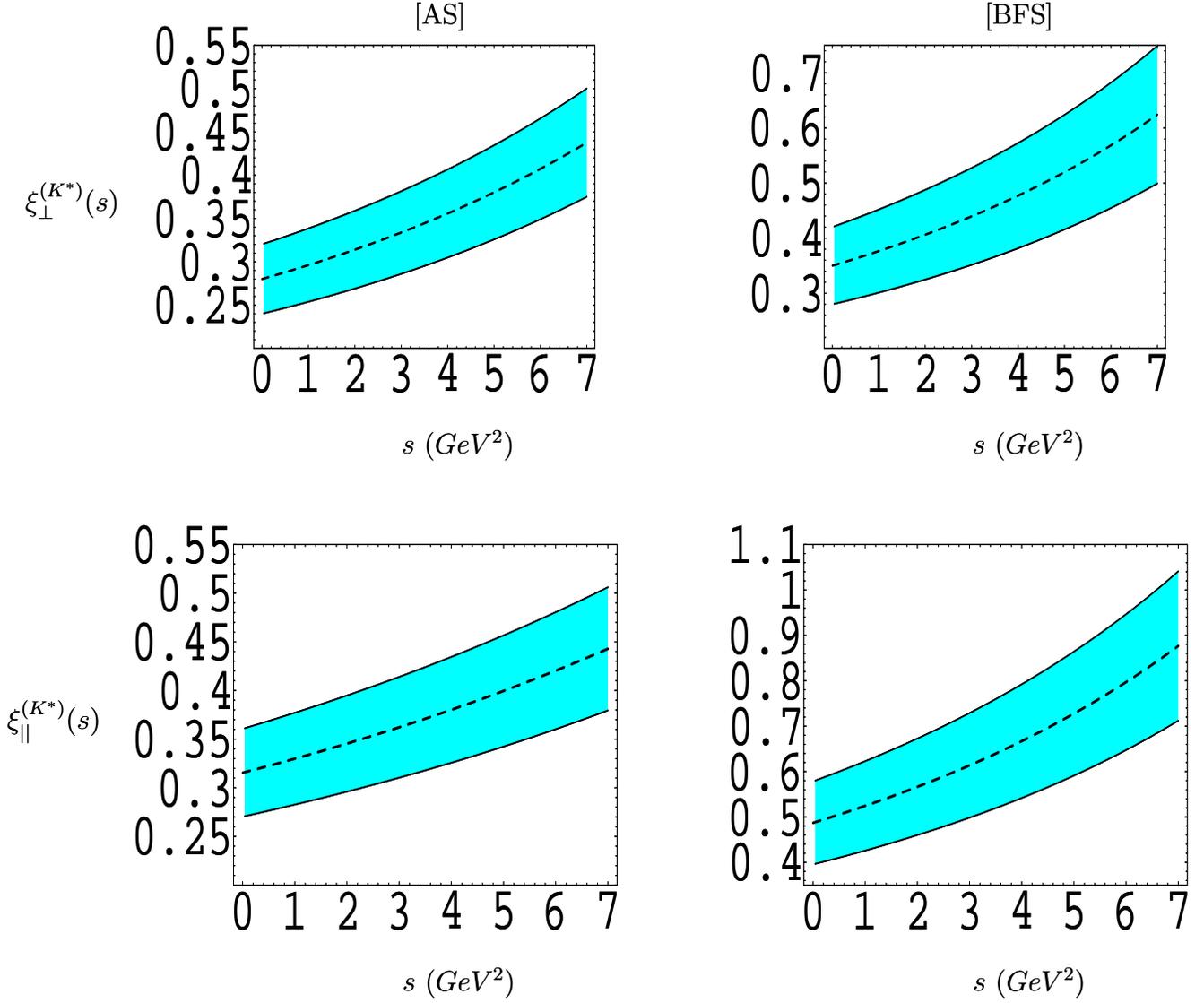}
\caption{LEET form factors $\xi^{(K^*)}_{{\perp},||}(s)$ 
for  $B \rightarrow  K^*\ l^+\ l^-$. 
The two columns denoted by [AS] and [BFS] represent, respectively, our
$\xi^{(K^*)}_{{\perp},||}(s)$ and the ones used by Beneke {\it et al.} in
ref\cite{Beneke:2001at}. The central values are represented by the dashed
curves, while the bands reflect the uncertainties on the form factors.}
\label{ffASetBF}
\end{center}
\end{figure}

\begin{table}[t] 
   \begin{center} 
     \begin{tabular}{| l l| l l |} 
\hline  
\rule[-2mm]{0mm}{7mm}
     $M_W$                           & $80.4$~GeV & 
        $f_B$                           & $200 \pm 20$~MeV\\ 
     
        $\hat m_t(\hat m_t)$            & $167 \pm 5$~GeV & 
         $f_{K^*,\parallel}$             & $225 \pm 30$~MeV \\

    $m_{b,pole}(2\,\mbox{GeV})$   & $4.6 \pm 0.1$~GeV & 
        $f_{K^*,\perp}(1\,\mbox{GeV})$  & $185 \pm 10$~MeV \\

         $m_c$                           & $1.4 \pm 0.2$~GeV &
       $f_{\rho}$~(1 GeV)             & $198\pm 7 $~MeV \\

      $\alpha_{em}$               & $ 1/137$ & 
        $\lambda_{B,+}^{-1}$            & $(3 \pm 1)$~GeV \\
        
        $\tau_B$                & 1.65 ps &
       $a_1( K^*)_{\perp,\,\parallel}$    & $0.2 \pm 0.1$ \\ 

      $|V_{ts}^* V_{tb}|$                      & $0.041 \pm 0.003$ &
       $a_2( K^*)_{\perp,\,\parallel}$    & $0.05 \pm 0.1$  \\

 $R_{b}=|V_{ub}|/|V_{ts}^* V_{tb}|$      & $0.094\pm 0.014$&
      $\xi^{(K^*)}_\perp(0)$                & $ 0.28 \pm 0.04$ \\

        $\Lambda_{\rm QCD}^{(n_f=5)}$   & $220 \pm 40$~MeV &
        $\xi^{(\rho)}_\perp(0)$                & $ 0.22 \pm 0.04$ \\
        
        $\langle \ell_{+}^{-1}\rangle^{(\rho)}_{+}$     & $0.3 \pm 0.2$~
        $(GeV)^{-1}$&
        $\langle \bar{u}^{-1}\rangle^{(\rho)}_{||}$            & 3.48 
        \\[0.15cm]
\hline
       \end{tabular} 
\end{center} 
\vspace{0.1cm}
\centerline{\parbox{14cm}{\caption{\label{parameters}
Input parameters and their uncertainties used in the 
calculations of the decay rates for $B \to K^* \ell^+ \ell^-$ and $B
\to \rho \ell \nu_\ell$ in the LEET approach.}}} \vspace{0.1cm}
\end{table} 
%
Lacking a complete solution of non-perturbative QCD, one has to rely on 
certain approximate methods to calculate the above form factors. In this 
paper, we take the ones given in \cite{Ali:1999mm}, obtained in the framework
of Light-cone QCD sum rules, and parametrized as follows:
\begin{equation}\label{eq:para}
F(\hat{s}) = F(0) \exp ( c_1 \hat{s} + c_2 \hat{s}^2 ).
\end{equation}
The coefficients in this parametrization are listed in Table \ref{tab:p1},
and the corresponding LEET form factors  $\xi_{\perp}(s)$ and
$\xi_{||}(s)$ are plotted in Fig.~\ref{ffASetBF}. The range
$\xi_{\perp}(s)=0.28\pm 0.04$ is determined by the $B \rightarrow \
K^{*}\gamma$ decay rate, calculated in the LEET approach in 
next-to-leading order \cite{Ali:2001ez,Bosch:2001gv,Beneke:2001at} and
current data. This gives somewhat smaller values for $T_1(0)$ and $T_2(0)$
than the ones estimated with the QCD sum rules.
%
%
\section{Distributions in the Decay $B \rightarrow \ K^{*} \ell^{+} 
\ell^{-}$}
\begin{figure}
\psfrag{a}{\hskip 0.3cm $s\ (GeV^2)$}
\psfrag{b}{\hskip - 1.5cm$ |H_{+}^{L}(s)|^2$}
\begin{center}
\includegraphics[width=12cm,height=9cm]{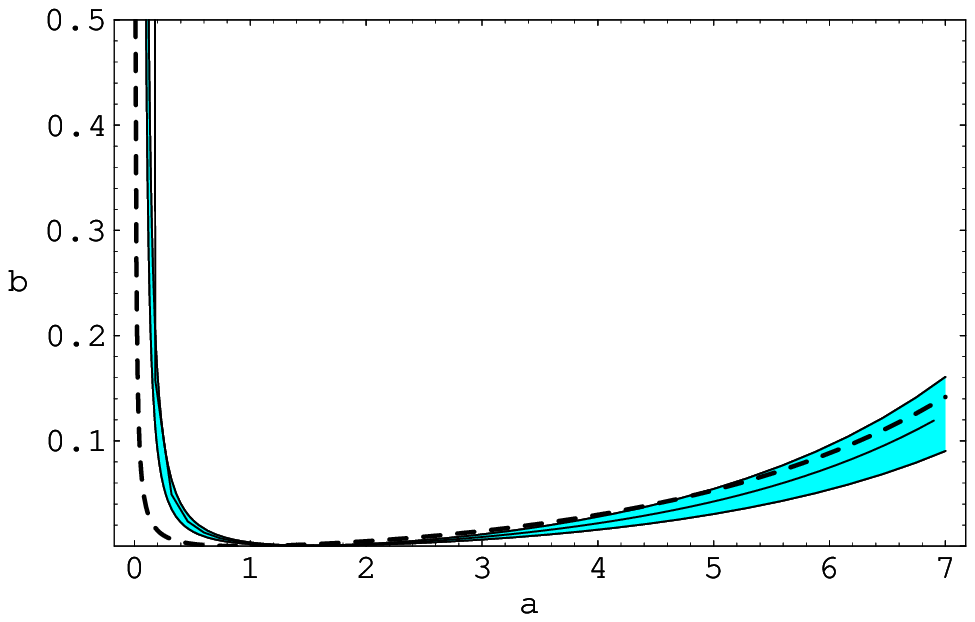}
\caption{ The helicity amplitude $ |H_{+}^{L}(s)|^2$ at next-to-leading
order (solid center line) and leading order (dashed). The band reflects
theoretical uncertainties from the input parameters.}
\label{figHpls1L2}
\end{center}
\end{figure}

\begin{figure}
\psfrag{a}{\hskip 0.3cm $s\ (GeV^2)$}
\psfrag{b}{\hskip - 1.5cm$ |H_{-}^{L}(s)|^2$}
\begin{center}
\includegraphics[width=12cm,height=9cm]{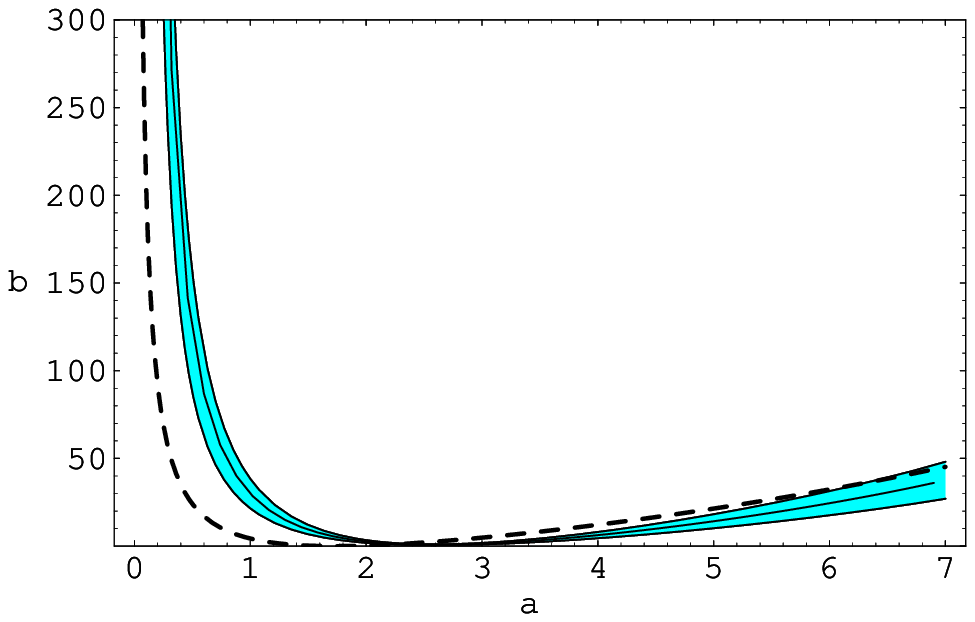}
\caption{The helicity amplitude $ |H_{-}^{L}(s)|^2$ at next-to-leading
order (solid center line) and leading order (dashed). The band reflects
theoretical uncertainties from the input parameters.}
\label{figHLmns12}
\end{center}
\end{figure}

\begin{figure}
\psfrag{a}{\hskip 0.3cm $s\ (GeV^2)$}
\psfrag{HRp}{\hskip - 1.5cm$ |H_{+}^{R}(s)|^2$}
\begin{center}
\includegraphics[width=12cm,height=9cm]{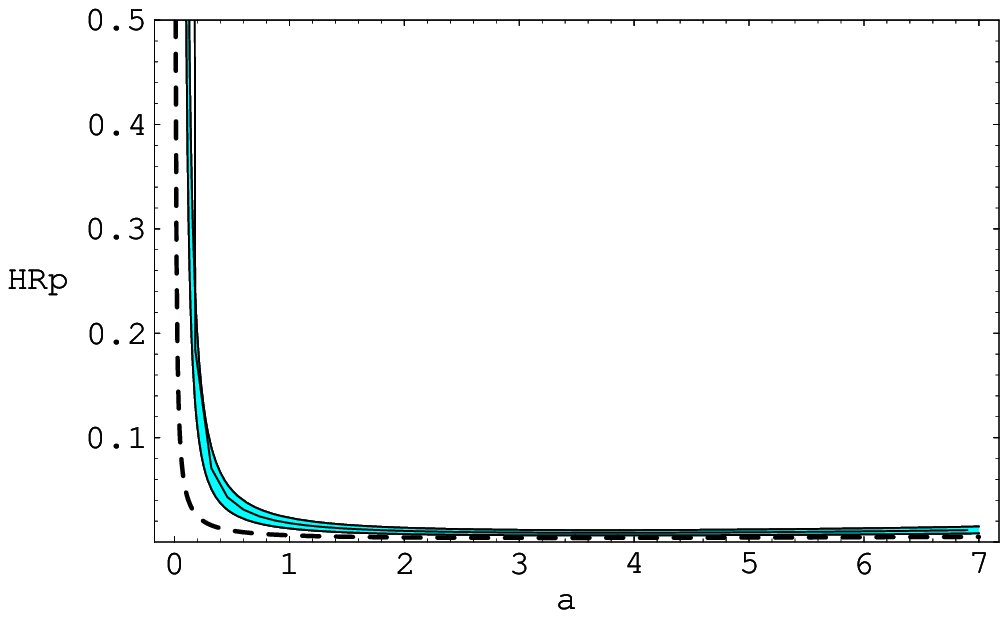}
\caption{ The helicity amplitude $ |H_{+}^{R}(s)|^2$ at next-to-leading
order (solid center line) and leading order (dashed). The band reflects
theoretical uncertainties from the input parameters.}
\label{figHRpls12}
\end{center}
\end{figure}

\begin{figure}
\psfrag{a}{\hskip 0.3cm $s\ (GeV^2)$}
\psfrag{HRm}{\hskip - 1.5cm$ |H_{-}^{R}(s)|^2$}
\begin{center}
\includegraphics[width=12cm,height=9cm]{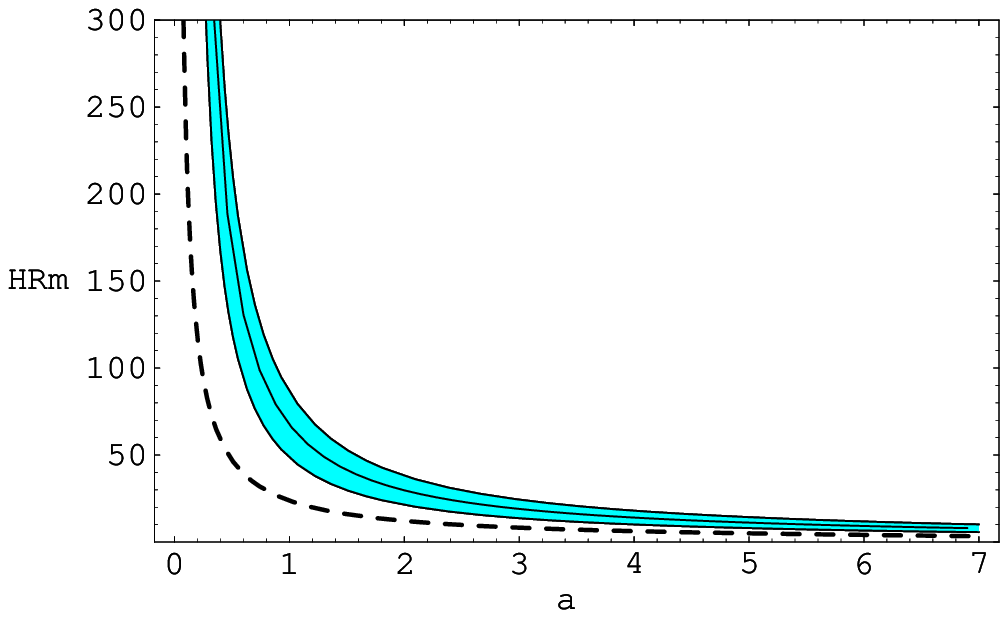}
\caption{ The helicity amplitude $ |H_{-}^{R}(s)|^2$ at next-to-leading
order (solid center line) and leading order (dashed). The band reflects
the theoretical uncertainties from the input parameters.}
\label{figHRmns12}
\end{center}
\end{figure}

\hspace*{\parindent}
We introduce the helicity amplitudes for the decay $B \rightarrow \
K^{*}(\rightarrow \ K(p_K) + \pi(p_\pi )) \ell^{+}(p_+) \ell^{-}(p_-)$,
which can be expressed as \cite{Kim:2000dq}:
\begin{eqnarray}
H_{\pm}^{L,R}(s)&=& (a_{L,R} \pm c_{L,R} \sqrt{\lambda})~,\nn \\
H_{0}^{L,R}(s) &=& -a_{L,R} {P.L\over m_{V} \sqrt{s}} +
{b_{L,R} \lambda \over m_{V} \sqrt{s}}~,\label{Hs}
\end{eqnarray}
\noi where $P.L = (m_B^2-m_V^2 - s)/2$ and $ \lambda=\left[{1\over 4}(m_{B}^2-m_{V}^2-s)^2 - m_{V}^2\
s\right]$ and $V$ stands here for the vector meson $K^*$. Our definitions for
the quantities $a_{L,R}$, $b_{L,R}$ and $c_{L,R}$ differ from those
used by Kim {\it et al.}~\cite{Kim:2000dq}
by a factor of $1/\sqrt{s}$. They read as follows:

\begin{eqnarray}
a_{L/R} &=&  {i(m_B +  m_V)\over {2\ m_b\ m_B \sqrt{s}}} \Big[s\ m_B
(\pm C_{10} - C_{9})  A_1(s) + 4 {\cal T}_1(s)\ m_b (m_V -  m_B) E_V\Big]
~,\label{aL}\\
b_{L/R} &=& {i\over m_b\ m_B\ (m_B^2 -m_V^2) \ \sqrt{s}} 
\Big[ 4\ {\cal T}_1(s) \ m_b \ (-m_B^2 + m_V^2) \ E_V 
\label{bL}\\
& & + m_B s \Big(-2\ m_b  \left\{ {\cal T}_1(s)+ {\cal
T}_3(s)-\frac{m_B}{2E_V}\,{\cal T}_2(s)\right \} 
+ A_2(s)\ (\pm C_{10} - C_{9})(m_B -m_V) \Big) \Big ]~,\nn\\
c_{L/R} &=&  {i\over {m_b\ (m_B + m_V)\ \sqrt{s}}} 
\Big[2 {\cal T}_1 (s)\ m_b \ (m_B + m_V) + (\mp C_{10} + C_9)\ s \
 V(s)\Big]~.\label{cL}
\end{eqnarray}
We show the helicity amplitudes $\vert H_{+}^{ L}(s) \vert^2$,
$\vert H_{-}^{L}(s) \vert^2$, $\vert H_{+}^{R}(s) \vert^2$,
and $\vert H_{-}^{R}(s) \vert^2$ in Fig.~\ref{figHpls1L2},
Fig.~\ref{figHLmns12}, Fig.~\ref{figHRpls12}, and Fig.~\ref{figHRmns12},
respectively.

\subsection{\bf Dalitz distributions}
\hspace*{\parindent}
Using the above helicity amplitudes, the angular distribution in  $B
\rightarrow K^{*}(\rightarrow K \pi )\ell^{+} \ell^{-}$ is given by the
following expression:
\begin{eqnarray}
\label{5diff}
&&{d^4\Gamma  \over  ds\ d\cos\theta_{K}\ d\cos\theta_{+}\ d\phi} = {3\
\alpha_{em}^2 G_{F}^2  \sqrt{\lambda}  m_{b}^2 |V_{tb}V_{ts}^*|^2 \over 128
(2\pi)^6 m_{B}^3 }\label{eq:angdist}  \\ 
&&\,\,\,\,\,\times \Big\{4\ \cos^2\theta_{K}\ \sin^2\theta_{+}\ 
\Big(|H^R_{0}(s)|^2+|H^L_{0}(s)|^2 \Big)\nn  \\ 
&&\,\,\,\,\, + \sin^2\theta_{K}\ (1+ \cos^2\theta_{+})\
\Big( |H^L_{+}(s)|^2+|H^L_{-}(s)|^2 +|H^R_{+}(s)|^2 +|H^R_{-}(s)|^2 \Big)\nn \\ 
&&\,\,\,\,\, - 2\  \sin^2\theta_{K}\  \sin^2\theta_{+}\ \Big[\
\cos2\phi \ {\rm Re}
\Big(H^R_{+}(s) H^{R*}_{-}(s)+H^L_{+}(s) H^{L*}_{-}(s)\Big)\nn \\ 
&&\,\,\,\,\, - \sin2\phi \ {\rm Im}\Big(H^R_{+}(s) H^{R*}_{-}(s) + H^L_{+}(s)
 H^{L*}_{-}(s)\Big)\ \Big]\nn \\
&&\,\,\,\,\,- \sin2\theta_{K}\ \sin2\theta_{+}\ \Big[\ \cos\phi\
 {\rm Re}\Big(H^R_{+}(s)H^{R*}_{0}(s) 
+H^R_{-}(s) H^{R*}_{0}(s)+H^L_{+}(s)H^{L*}_{0}(s)+H^L_{-}(s) H^{L*}_{0}(s)
 \Big)\nn \\
&&\,\,\,\,\, -\sin\phi\ {\rm Im} \Big(H^R_{+}(s) H^{R*}_{0}(s) - H^R_{-}(s)
 H^{R*}_{0}(s)
+ H^L_{+}(s) H^{L*}_{0}(s) -H^L_{-}(s)H^{L*}_{0}(s)\Big)\ \Big]\nn \\
&&\,\,\,\,\, -2\ \sin^2\theta_{K}\  \cos\theta_{+}\
\Big(|H^R_{+}(s)|^2-|H^R_{-}(s)|^2 -|H^L_{+}(s)|^2+ |H^L_{-}(s)|^2 \Big)\nn \\
&&\,\,\,\,\, +2\ \sin\theta_{+}\  \sin2\theta_{K}\ 
\Big[\ \cos\phi\ {\rm Re}\Big(H^R_{+}(s) H^{R*}_{0}(s)-
H^R_{-}(s)H^{R*}_{0}(s)
- H^L_{+}(s) H^{L*}_{0}(s) + H^L_{-}(s)H^{L*}_{0}(s) \Big)\nn \\
&&\,\,\,\,\, - \sin\phi\ {\rm Im} \Big(  H^R_{+}(s) H^{R*}_{0}(s) +
H^R_{-}(s)
 H^{R*}_{0}(s) 
- H^L_{+}(s) H^{L*}_{0}(s)- H^{L}_{-}(s) H^{L*}_{0}(s)  \Big) \Big]
\Big\}~.\nn
\end{eqnarray}
\noi Here, the various angles are defined as follows: $\theta_{K}$ is the
polar angle of the K meson  in the rest system of the $K^*$
meson, measured with respect to the helicity axis, {\it{i.e.}}, the outgoing
direction of the $K^*$. Similarly, $\theta_{+}$ is the polar angle of the
positively charged lepton $\ell^+$ in the
dilepton rest system, measured with respect to the helicity axis of the
dilepton, and  $\phi$ is the azimuthal angle between the two planes 
defined by the momenta of the decay products
 $K^* \rightarrow K  \pi$ and $\gamma^* \rightarrow\ \ell^{+} \ell^{-}$.

\begin{figure}
\psfrag{c}{\hskip 0.3cm $s\ (GeV^2)$}
\psfrag{a}{\hskip -4.cm $d^2{\cal B}_{|H_{-}|^2}/ ds\ d\cos\theta_{+}\ 10^{-8}$}
\psfrag{b}{$ \cos\theta_{+}$}
\begin{center}
\includegraphics[width=12cm,height=9cm]{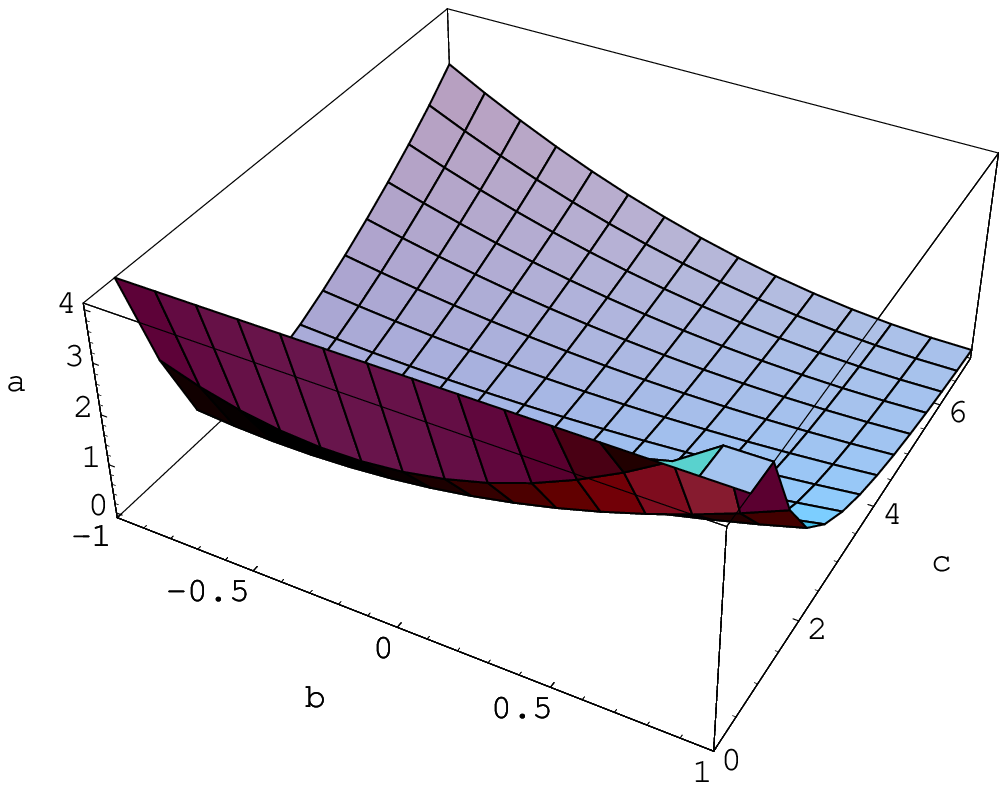}
\caption{Partial Dalitz distribution ${d^2{\cal B}_{|H_{-}|^2}
 (B \rightarrow  K^*\ l^+ \ l^-)\over d\cos\theta_{+}\ ds}$ .}  
 \label{dalitzdBrH-1dcostheta+}
\psfrag{c}{\hskip 0.3cm $s\ (GeV^2)$}
\psfrag{a}{\hskip -4.3cm $d^2{\cal B}_{|H_{0}|^2}/ ds\ d\cos\theta_{+}\ 10^{-8}$}
\psfrag{b}{$ \cos\theta_{+}$}
\includegraphics[width=12cm,height=9cm]{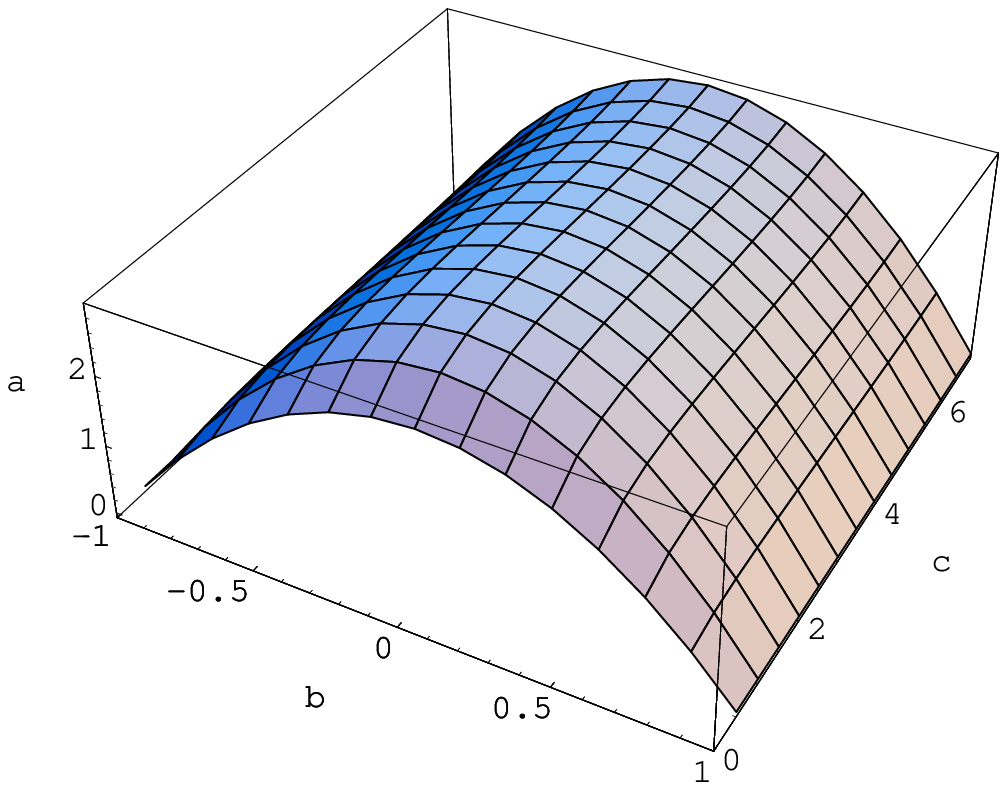}
\caption{Partial Dalitz distribution ${d^2{\cal B}_{|H_{0}|^2}(B \rightarrow 
 K^*\ l^+ \ l^-) \over 
d\cos\theta_{+}\ ds}$.}
\label{dalitzdBrH0dcostheta+}
\end{center}
\end{figure}

\begin{figure}
\psfrag{c}{\hskip 0.3cm $s\ (GeV^2)$}
\psfrag{a}{\hskip -3.5cm $d^2{\cal B}/ ds\ d\cos\theta_{+}\ 10^{-8}$}
\psfrag{b}{$ \cos\theta_{+}$}
\begin{center}
\includegraphics[width=12cm,height=9cm]{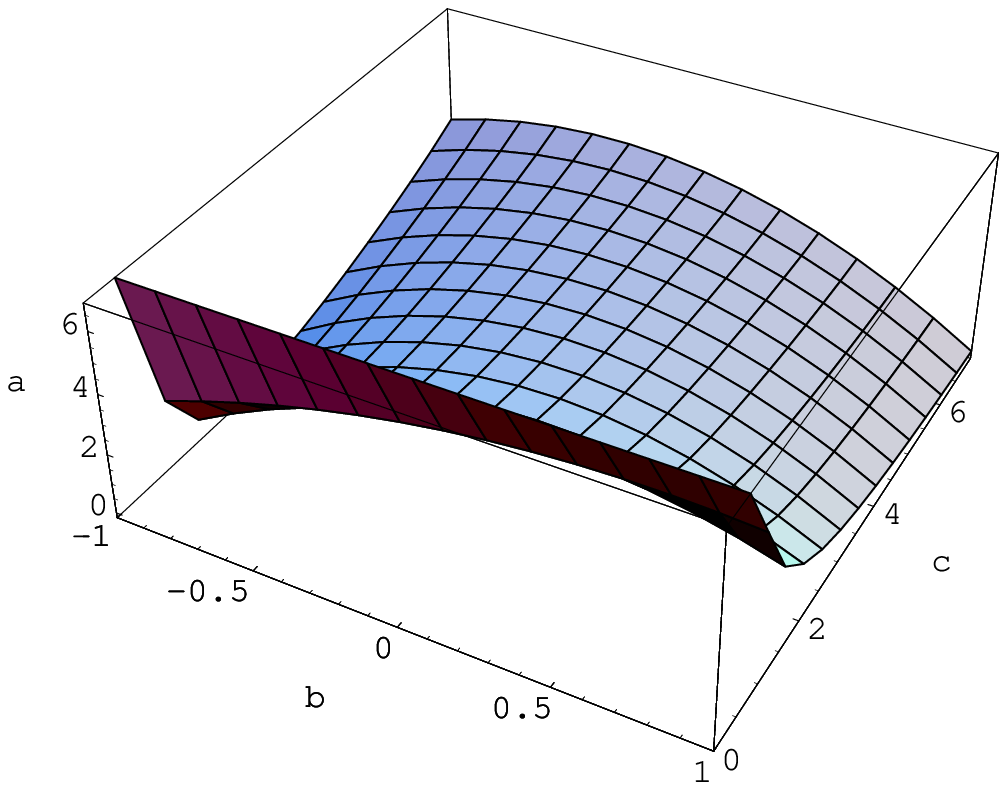}
\caption{Dalitz distribution ${d^2{\cal B}(B \rightarrow  K^*\ l^+ \ l^-) \over d\cos\theta_{+}\ ds}$.}
\label{dalitzdBrdcostheta+}
\end{center}
\end{figure}

Integrating over the angle  $\theta_{K}$ and $\theta_{+}$, we get the
Dalitz distribution in the remaining two variables  ($\phi,s)$:
\begin{eqnarray} 
\label{dBrdl2dphi}
&&{d^2{\cal B} \over d\phi\ ds} = \tau_B {\alpha_{em}^2 G_{F}^2\over 384 \pi^5}
\sqrt{\lambda} {m_{b}^2\over m_{B}^3 }  |V_{tb}V_{ts}^*|^2 \ {1\over
2\pi}\ \Big\{ |H_{0}(s)|^2 + |H_{+}(s)|^2+ |H_{-}(s)|^2 \\ 
&& -  \cos2\phi \ {\rm Re}\Big(H_{+}^R(s) H_{-}^{R*}(s)+H_{+}^L(s)
H_{-}^{L*}(s)\Big) + \sin2\phi \
 {\rm Im} \Big(H_{+}^R(s) H_{-}^{R*}(s)+H_{+}^L(s) H_{-}^{L*}(s)\Big)\Big\}
~.\nn
\end{eqnarray}
\noi where  $\tau_B$ is the $B$-meson life time, and~:
\begin{eqnarray} 
|H_{0}(s)|^2 & = &|H^L_{0}(s)|^2+|H^R_{0}(s)|^2~,\nn\\
|H_{+}(s)|^2 &=&  |H^L_{+}(s)|^2+|H^R_{+}(s)|^2~,\nn\\
|H_{-}(s)|^2 &=& |H^L_{-}(s)|^2 +|H^R_{-}(s)|^2~.\label{H2def}
\end{eqnarray}
\noi Similarly, we can get the Dalitz distributions in ($\theta_{K},s)$ and
$(\theta_{+},s)$, which read as follows:
\begin{eqnarray} 
\label{dBrdl2dcostetaK}
{d^2{\cal B} \over d\cos\theta_{K}\ ds} &=& \tau_B {\alpha_{em}^2 G_{F}^2\over 384
\pi^5}  \sqrt{\lambda} {m_{b}^2\over m_{B}^3 }  |V_{tb}V_{ts}^*|^2 \ {3\over 4}\
 \Big\{ 2\ \cos^2\theta_{K} \ |H_{0}(s)|^2\nn \\  
&& + \sin^2\theta_{K}\ \Big( |H_{+}(s)|^2+ |H_{-}(s)|^2\Big) \Big\}~. 
\end{eqnarray}
\begin{eqnarray} 
\label{dBrdl2dcosteta+}
{d^2{\cal B} \over d\cos\theta_{+}\ ds} &=& \tau_B {\alpha_{em}^2 G_{F}^2\over 384
\pi^5}  \sqrt{\lambda} {m_{b}^2\over m_{B}^3 }  |V_{tb}V_{ts}^*|^2 \ {3\over 8}\ \Big\{ 2\ \sin^2\theta_{+} \ |H_{0}(s)|^2  \\ 
&& +(1+ \cos\theta_{+})^2\ |H_{+}^L(s)|^2 +(1- \cos\theta_{+})^2\ |H_{+}^R(s)|^2\nn\\
&&+(1- \cos\theta_{+})^2\ |H_{-}^L(s)|^2 +(1+ \cos\theta_{+})^2\ |H_{-}^R(s)|^2 \Big\} \nn \\
&& ={d^2{\cal B}_{|H_{0}|^2} \over d\cos\theta_{+}\ ds}+ {d^2{\cal B}_{|H_{-}|^2} \over
d\cos\theta_{+}\ ds}+{d^2{\cal B}_{|H_{+}|^2} \over d\cos\theta_{+}\
ds}~.\label{dBrdl2dcosteta+Hs}
\end{eqnarray} 
In Figs.~(\ref{dalitzdBrH-1dcostheta+}),
(\ref{dalitzdBrH0dcostheta+}) and (\ref{dalitzdBrdcostheta+}), we plot,
respectively, the Dalitz distribution given by the two dominant partial
contributions and the complete expression given in
Eq.~(\ref{dBrdl2dcosteta+}).
%

%
%
\subsection{\bf Dilepton invariant mass spectrum}
\begin{figure}
\psfrag{a}{$s\ (GeV^2)$}
\psfrag{b}{\hskip -3.cm $d{\cal B}_{|H_{-}|^2}/ ds\ 10^{-7}$}
\psfrag{c}{}
\begin{center}
\includegraphics[width=12cm,height=9cm]{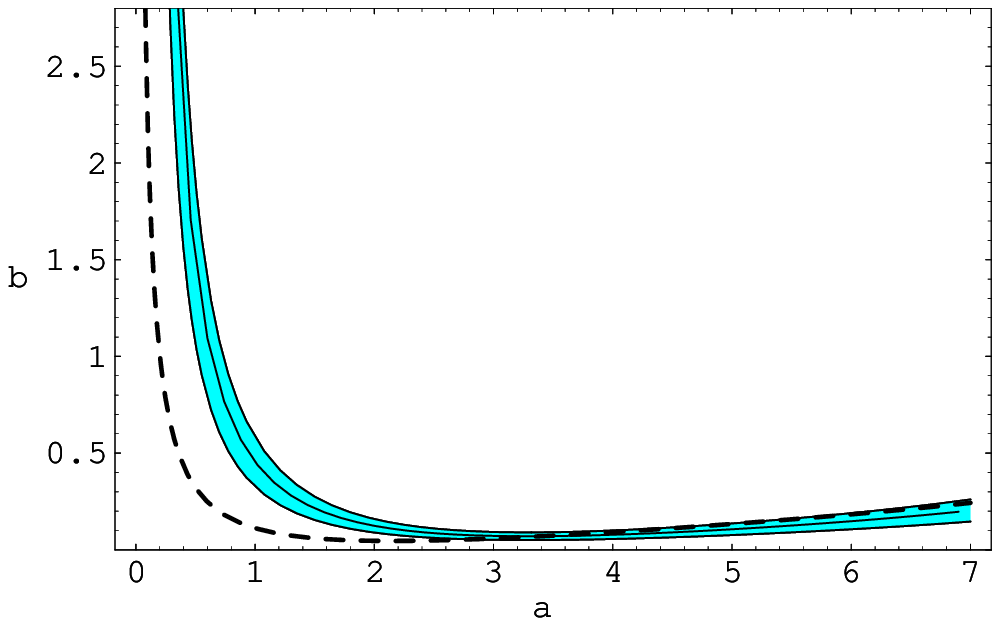}
\caption{The dilepton invariant mass distribution
$d{\cal B}_{|H_{-}|^2}/ ds$ for $ B \rightarrow
K^* \ell^+ \ell^-$ at next-to-leading order (solid center line) and
leading order
(dashed). The band reflects the theoretical uncertainties from input
parameters.}
\label{figdBrH-1}
\psfrag{a}{$s\ (GeV^2)$}
\psfrag{b}{\hskip -3.cm $d{\cal B}_{|H_{0}|^2}/ ds\ 10^{-7}$}
\psfrag{c}{}
\includegraphics[width=12cm,height=9cm]{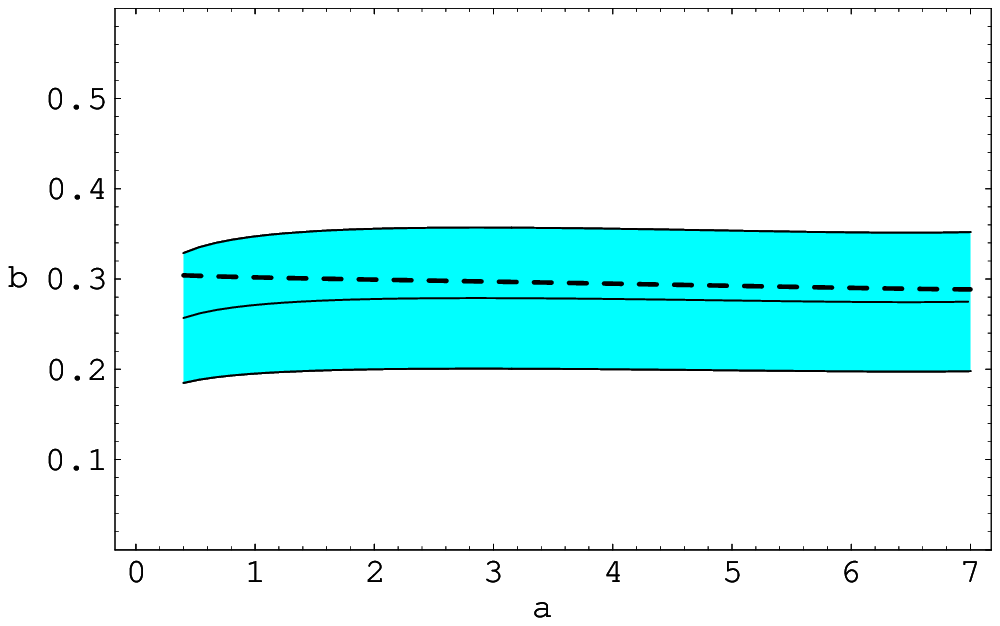}
\caption{The dilepton invariant mass distribution
$d{\cal B}_{|H_{0}|^2}/ ds$ for $ B \rightarrow
K^* \ell^+ \ell^-$ at next-to-leading order (solid center line) and
leading order
(dashed). The band reflects theoretical uncertainties from input
parameters.}
\label{figdBr0}
\end{center}
\end{figure}

\begin{figure}
\psfrag{a}{$s\ (GeV^2)$}
\psfrag{b}{\hskip -2.5cm $d{\cal B}/ ds\ 10^{-7}$}
\psfrag{c}{}
\begin{center}
\includegraphics[width=12cm,height=9cm]{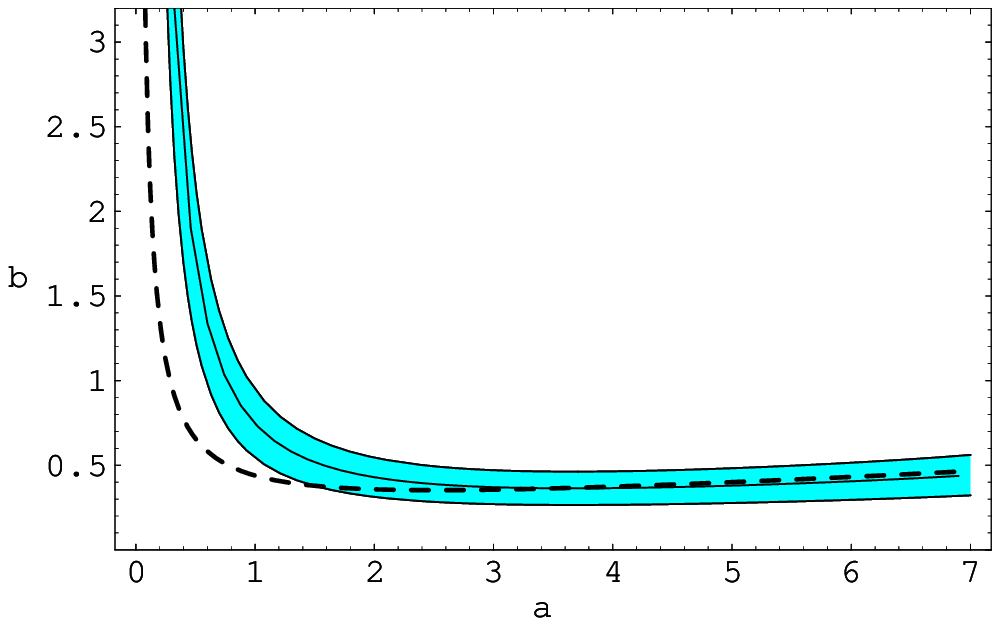}
\caption{The dilepton invariant mass distribution for $ B \rightarrow
K^* \ell^+ \ell^-$ at next-to-leading order (solid center line) and
leading
order (dashed). The band reflects theoretical uncertainties from
the input parameters.}
\label{figdBr}
\end{center}
\end{figure}

\hspace*{\parindent}
The dilepton invariant mass spectrum can be obtained by  integrating over
the angle variables, yielding:
\begin{eqnarray} 
\label{dBrdl2}
{d{\cal B} \over ds} &=& \tau_B {\alpha_{em}^2 G_{F}^2\over 384 \pi^5} 
 \sqrt{\lambda} {m_{b}^2\over m_{B}^3 }  |V_{tb}V_{ts}^*|^2 \Big\{|H_{+}(s)|^2 +
 |H_{-}(s)|^2+|H_{0}(s)|^2 \Big\}~. 
\end{eqnarray}
In LEET, the helicity amplitudes (\ref{Hs}) are expressed as:
\begin{eqnarray}
H^{L/R}_{+}(s) &=& {i\over\ {2\ m_b\ m_B\ (m_B + m_V) \sqrt{s}}}
\Big[ -4\ {\cal T}_1(s)\ m_b\ (m_B - m_V)\ (m_B + m_V)^2\ E_V\nn\\
&& +(\pm C_{10}-C_{9})\ m_B\ (m_B + m_V)^2 \  s \ A_1(s)\nn\\ 
&& +2 m_B \sqrt{\lambda} \Big \{ 2 {\cal T}_1(s)\ m_b\ (m_B+ m_V) +
(\mp C_{10}+C_{9})\ s\ V(s) \} \Big \} \Big] ~,\label{HK+}
\end{eqnarray}
\begin{eqnarray}
H^{L/R}_{-}(s) &=& {i\over\ {2\ m_b\ m_B\ (m_B + m_V)}\sqrt{s}} 
\Big[ -4\ {\cal T}_1(s)\ m_b\ (m_B - m_V)\ (m_B + m_V)^2\ E_V\nn\\
&& +(\pm C_{10} - C_{9})m_B (m_B + m_V)^2\  s \ A_1(s)\nn\\ 
&& -2 m_B \sqrt{\lambda} \Big \{2 {\cal T}_1(s)\ m_b (m_B + m_V)
+ (\mp C_{10} + C_{9})s V(s) \} \Big \} \Big]~,\label{HK-} 
\end{eqnarray}
%
\begin{eqnarray}
H^{L/R}_{0}(s)&=& {i\over {4\ m_b \ m_B \ m_V (-m_B^2 + m_V^2)s}}
\Big[ 8\  \lambda\  m_b\ {\cal T}_1(s) \Big \{2 (m_B^2 - m_V^2) E_V + m_B\ s\Big \}\nn\\
&& + 4\ \lambda\  m_B\ s \Big \{ 2\  m_b\ ({\cal T}_3(s)-\frac{m_B}{2\
E_V}\,{\cal T}_2(s))-A_2(s)(\pm C_{10} - C_{9}) (m_B - m_V)\Big \}\nn\\
&& +(m_B - m_V) (m_B + m_V)^2 (m_B^2 - m_V^2 -s)\Big \{4\ {\cal T}_1(s)\
m_b\ E_V  (-m_B + m_V)\nn\\
&&+ s\  m_B\ A_1(s)(\pm C_{10} - C_{9})\Big \} \Big]~. \label{HK0}
\end{eqnarray}
\noi In Figs.~(\ref{figdBrH-1}),  (\ref{figdBr0}) and
(\ref{figdBr}) we have plotted, respectively, the dilepton invariant
mass spectrum $d{\cal
B}_{|H_{-}|^2}/ ds$, $d{\cal B}_{|H_{0}|^2}/ ds$ and the total dilepton
invariant mass, showing in each case the leading order and the 
next-to-leading order results. The contribution
proportional to the helicity amplitude $H_{+}(s)$ is negligible, and hence
not shown, but is is included in calculating the total dilepton spectrum. As
can be seen from
Figs.~(\ref{figdBrH-1}) and (\ref{figdBr}) the total decay rate is
dominated by the contribution from the helicity $|H_{-}|$ component. 
The next-to-leading order correction to the lepton invariant mass
spectrum in $ B \rightarrow K^* \ell^+ \ell^-$ is significant in the 
low dilepton mass region ($s \leq 2$ GeV$^2$), but small beyond that
shown for the anticipated validity of the LEET theory ($s \leq 8$
GeV$^2$). Theoretical uncertainty in our prediction is mainly due to
the form factors, and to a lesser extent due to the
parameters $\lambda_{B,+}^{-1}$ and the $B$-decay
constant, $f_B$.
%

\subsection{\bf Forward-Backward asymmetry}
\hspace*{\parindent}
The differential forward-backward asymmetry (FBA) is defined as
 \cite{Ali:1991is}
\begin{equation}
  \frac{\d \a_{\rm FB}}{\d \sh} \equiv 
        -\int_0^{\uh(\sh)} \d\uh \frac{\d^2\gl}{\d\uh \d\sh}
              + \int_{-\uh(\sh)}^0 \d\uh \frac{\d^2\gl}{\d\uh \d\sh} \; .
  \label{eq:dfba}
\end{equation}
The kinematic variables $(\hat{s},\hat{u})$ are defined as follows
\begin{eqnarray}
  \hat{s} & \equiv & {q^2\over m_{B}^2}  \; , \\
  \hat{u} & \equiv & (\hat{p}_{B} - \hat{p}_-)^2 - (\hat{p}_{B} -
\hat{p}_+)^2 \ ,; 
\end{eqnarray}
which are bounded as 
\begin{eqnarray}
  (2 \hat{m}_{l})^2 \leq & \hat{s} & \leq (1 - \hat{m}_{K^*})^2  \; ,
  \label{eq:sbound}\\
  -\hat{u}(\hat{s}) \leq & \hat{u} & \leq \hat{u}(\hat{s}) \; ,
  \label{eq:ubound}
\end{eqnarray}
with $\hat{m}_{\ell}=m_{\ell}/m_{B}$, and
\begin{eqnarray}
 \hat{u}(\hat{s}) = {2\over m_{B}^2} \sqrt{\lambda (1-4 {\hat{m}_{l}^2\over
 \hat{s}})}~. 
\end{eqnarray}
Note that the variable $\hat{u}$ corresponds to $\theta_{+}$, the angle 
between the momentum of the $B$-meson and the positively charged lepton 
$\ell^+$  in the dilepton CMS frame through the relation
$\hat{u} = -\hat{u}(\hat{s}) \cos \theta_{+}$ \cite{Ali:1991is}.

\begin{figure}
\psfrag{a}{$s\ (GeV^2)$}
\psfrag{b}{\hskip -2.5cm $dA_{FB}/ ds$}
\psfrag{c}{\hskip 0cm }
\begin{center}
\includegraphics[width=12cm,height=9cm]{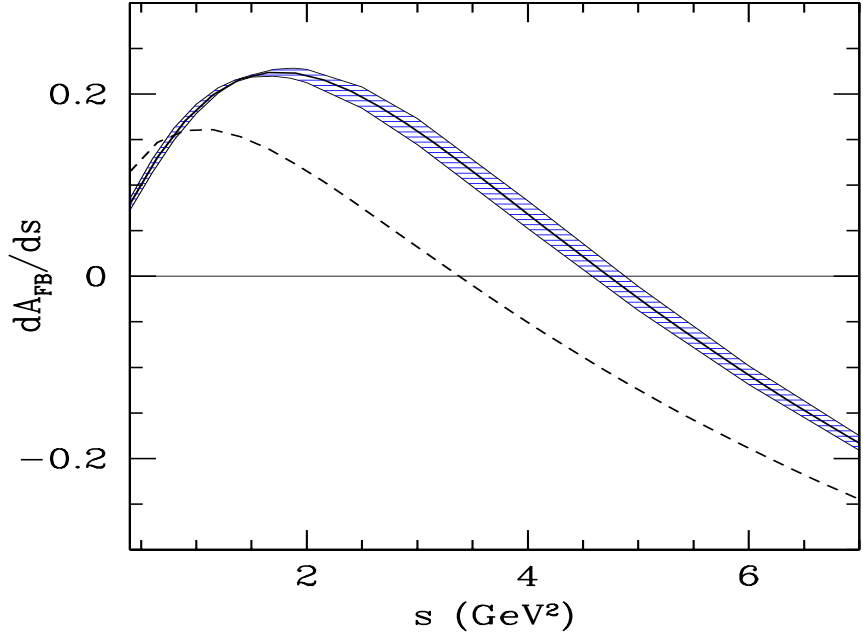}
\caption{Forward-backward asymmetry $dA_{FB}(B \rightarrow K^* l^+
l^-)/ds$ at next-to-leading order (solid center line) and leading
order (dashed). The band reflects the theoretical uncertainties from
the input parameters.}
\label{FigFBA}
\end{center}
\end{figure}

At the leading order, the FBA  in $B\to K^*\ell^+\ell^-$ decays reads as follows
\begin{eqnarray}
  \frac{\d \a_{\rm FB}}{\d \sh}& =& 
\frac{G_F^2 \, \alpha_{em}^2 \, m_B^5}{2^{8} \pi^5} 
      \left| V_{ts}^\ast  V_{tb} \right|^2 \, \sh \uh(\sh)^2
 \label{eq:dfbabvllex} \\
& & \times  \ct 
\left[ (-\cse) {\mbh\over\ \sh}(-1 + \mvh^2 + \sh) + 
2\ {E_V\over m_B}(\cse\ {\mbh\over\ \sh} +  {\rm Re}[\cne]) \right]
 \xi_{\perp}(s)^2 \nn \; .
\end{eqnarray}
The position of the zero of this function, $\sh_0$,  is given by
solving the following equation:
\begin{eqnarray}
{\rm Re}(\cne(\sh_0)) =- \frac{\mbh}{\sh_0} \cse 
\left\{\frac{1-\mvh^2-\sh}{1+\mvh^2-\sh} +1 \right\} \; .
\label{eq:fbzero}
\end{eqnarray}
Our results for FBA are shown in Fig.~\ref{FigFBA} in the  LO and NLO
accuracy. We essentially confirm the results obtained in the NLO-LEET
context by Beneke {\it et al.} \cite{Beneke:2001at}.

%
\section{Transversity Amplitudes for $B \to K^* \ell^+ \ell^-$ 
and Comparison with Data on $B \to K^* J/\psi (\to \ell^+ \ell^-)$}
\begin{figure}[!hbtp]
\begin{center}
\epsfxsize8.6cm
\includegraphics[width=12cm,height=9cm]{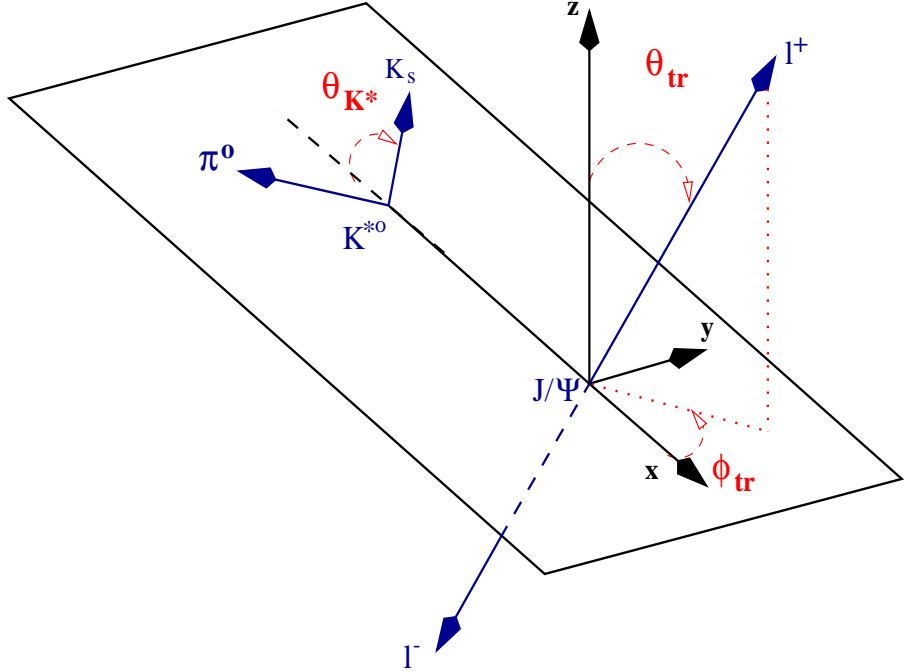}
\caption{Definitions of the transversity angles \thetatr , \phitr, and
\thetakstar . The angles \thetatr~and \phitr~are determined in the
\jpsi\ rest frame. The angle  \thetakstar~is determined in the $K^*$ rest
 frame.}
\label{fig:jpsiksttrans}
\end{center}
\end{figure}

\hspace*{\parindent}
The decay $B \to J/\psi K^*$ is described
by three amplitudes $({\cal A}_i; i=0,\parallel, \perp)$ in the
transversity basis, where \az, \ap\ and
\at\ have CP eigenvalues $+1, +1$ and $-1$, respectively
\cite{ref:dunietz,ref:dighe}, and should not be confused with the form
factors $A_0(s)$, $A_1(s)$ etc. Here,
\az\ corresponds to the longitudinal polarization of the vector meson
$K^*$ and \ap\ and \at\ correspond to parallel and 
transverse polarizations, respectively. 
The relative phase between the parallel (transverse)
amplitude and the longitudinal amplitude is given by
$\phi_\parallel(s)\equiv\mbox{arg}\Big( \ap/\az  \Big)$ 
$\Big(\phi_\perp(s)\equiv\mbox{arg}\Big(\at/\az \Big)\Big)$. The
transversity
frame is defined as the \jpsi\ rest frame (see Fig.~\ref{fig:jpsiksttrans}).
The $K^*$ direction defines the negative $x$ axis. The $K\pi$ decay plane
defines the $(x,y)$ plane, with $y$ oriented such that $p_y(K) > 0$. The
$z$ axis is the normal to this plane, and the coordinate system is
right-handed.  The transversity angles 
\thetatr\ and \phitr\ are defined as the polar and azimuthal angles of the
positively charged  lepton from the \jpsi\ decay; \thetakstar\ is the $K^*$
helicity
angle defined in the $K^*$ rest frame as the angle between the $K$
direction and the direction opposite to the \jpsi. This basis
has been used by the CLEO \cite{ref:cleo},
CDF \cite{ref:cdf}, BABAR \cite{ref:babar}, and the BELLE \cite{ref:belle}
collaborations to project out the amplitudes in the decay $B \to J/\psi
K^*$ with well-defined CP eigenvalues in their measurements of the 
quantity $\sin 2 \beta$, where $\beta$ is an inner angle of the unitarity
triangle. We also adopt this basis and analyze the various amplitudes
from the non-resonant (equivalently short-distance) decay $B \to K^*
\ell^+ \ell^-$. In this basis, both the
resonant $B \to K^* J/\psi \to K^* \ell^+ \ell^-$ (already measured) and
the non-resonant ($B \to K^* \ell^+ \ell^-$) amplitudes turn out to be
very similar, as we show here.

The angular distribution is given in terms of the linear
polarization basis (${\cal A}_{\pm1}(s)=({\cal A}_{\parallel}(s)\pm
{\cal A}_{\perp}(s))/\sqrt{2}$) and ${\cal A}_0(s)$ by
\begin{eqnarray}
\label{eqn:distrib}
\gfrac{\dd^{4} \Gamma }{ds\;\,\dd\cthetatr\;\,\dd\cthetakstar\;\,\dd\phitr} 
& = & f_1(w)\cdot\azd  + f_2(w)\cdot\apd  + f_3(w)\cdot\atd    \nonumber \\
& + & \eta f_4(w)\cdot\pipt + f_5(w)\cdot\przp  \nonumber\\
& + &  \eta f_6(w)\cdot\pizt~,   \nonumber
\end{eqnarray}
\noi where $\eta=+1(-1)$ for $B^0$ and $B^+$ ($\bar{B}^0$ and $B^-$),
and the coefficients $f_{i=1,...,6}$, which depend on
the transversity angles $w=(\thetakstar, \thetatr, \phitr)$, are given by:
\begin{eqnarray*}
f_1(w) & = & ~ ~ \ \, 9/(32\pi)\cdot\, 
 2\cq{\thetakstar}(1-\sq{\thetatr}\cq{\phitr}), \\
f_2(w) & = & ~ ~ \ \, 9/(32\pi)\cdot\,
 \sq{\thetakstar}(1-\sq{\thetatr}\sq{\phitr}), \\
f_3(w) & = & ~ ~ \ \, 9/(32\pi)\cdot\,
 \sq{\thetakstar}\sq{\thetatr}, \\
f_4(w) & = & ~ ~ \ \, 9/(32\pi)\cdot\,
 \sq{\thetakstar}\sin{2\thetatr}\sphitr,\\
f_5(w) & = & - \ \, 9/(32\pi)\cdot\,
  1/\sqrt{2}\cdot\sin{2\thetakstar}\sq{\thetatr}\sin{2\phitr}, \\
f_6(w) & = & ~ ~ \ \, 9/(32\pi)\cdot\,
  1/\sqrt{2}\cdot\sin{2\thetakstar}\sin{2\thetatr}\cos{\phitr}~.
\end{eqnarray*}
\noi In terms of the helicity amplitudes
$H^{L/R}_{\pm 1,0}$, introduced earlier, the amplitudes in the linear
polarization basis, ${\cal A}_{0,\perp,\parallel}$, can be calculated  
from the relation:
\begin{eqnarray*}
{\cal A}_0(s)  &=& \kappa \Big(H^L_{0}(s)+H^R_{0}(s)\Big )~,\\
{\cal A}_{+1}(s)&=&\kappa \Big(H^L_{+}(s)+H^R_{+}(s)\Big)~,\\
{\cal A}_{-1}(s)&=&\kappa \Big(H^L_{-}(s)+H^R_{-}(s)\Big)~,
\end{eqnarray*}
with
$\kappa^2= {\alpha_{em}^2 G_{F}^2\over 384 \pi^5}
\sqrt{\lambda} {m_{b}^2\over m_{B}^3 }  |V_{tb}V_{ts}^*|^2~$.
\par
\begin{figure}[!hbtp]
\begin{center}
\includegraphics[width=12cm,height=9cm]{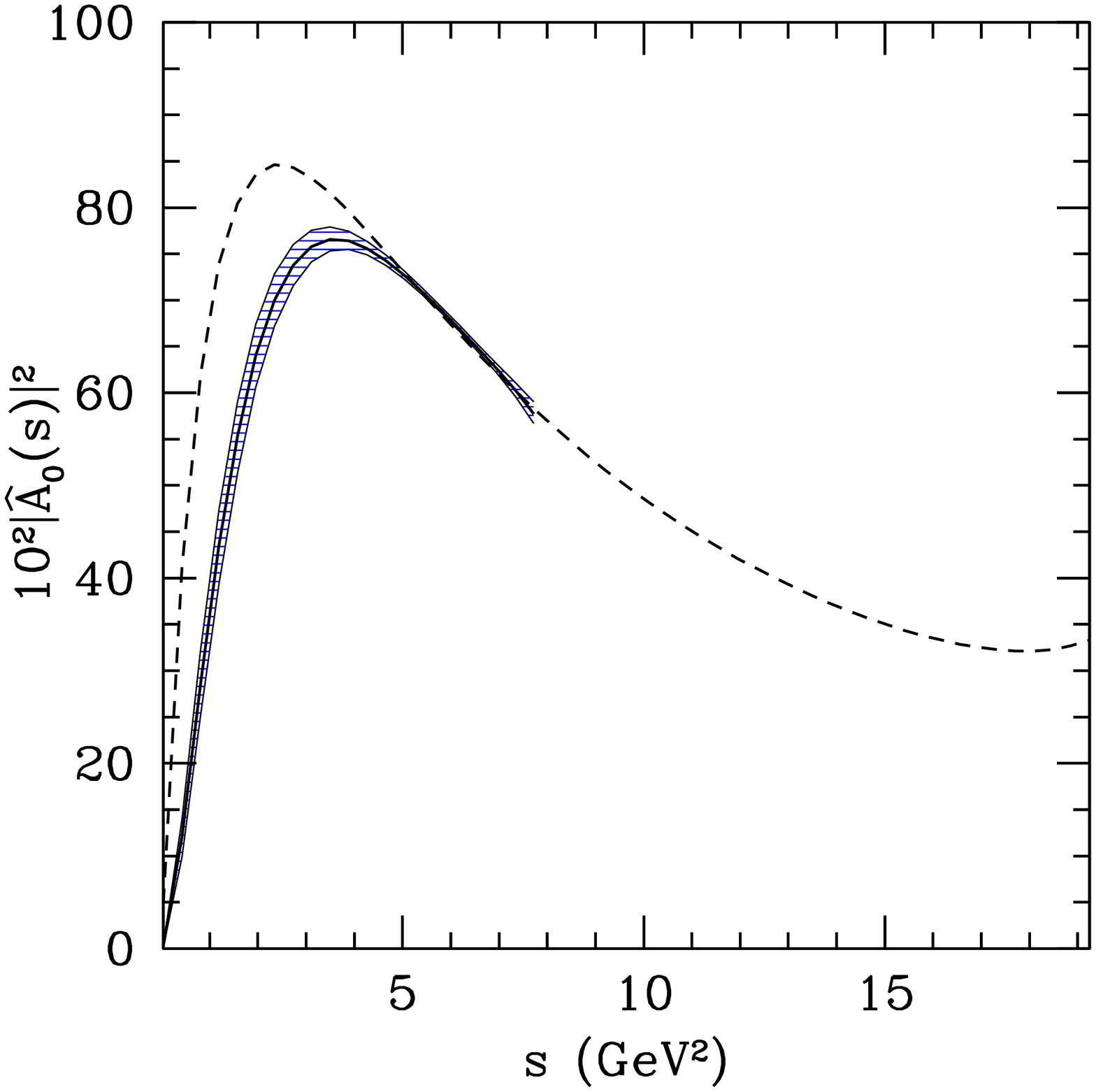}
\caption{ The helicity amplitude $|\hat {\cal A}_0(s)|^2$ in
 $B \to K^* \ell^+ \ell^-$ at next-to-leading order
(center line) and leading order (dashed). The band for NLO reflects 
theoretical uncertainties from input parameters.}
\label{figA02}
\end{center}
\end{figure}
\begin{figure}[!hbtp]
\begin{center}
\includegraphics[width=12cm,height=9cm]{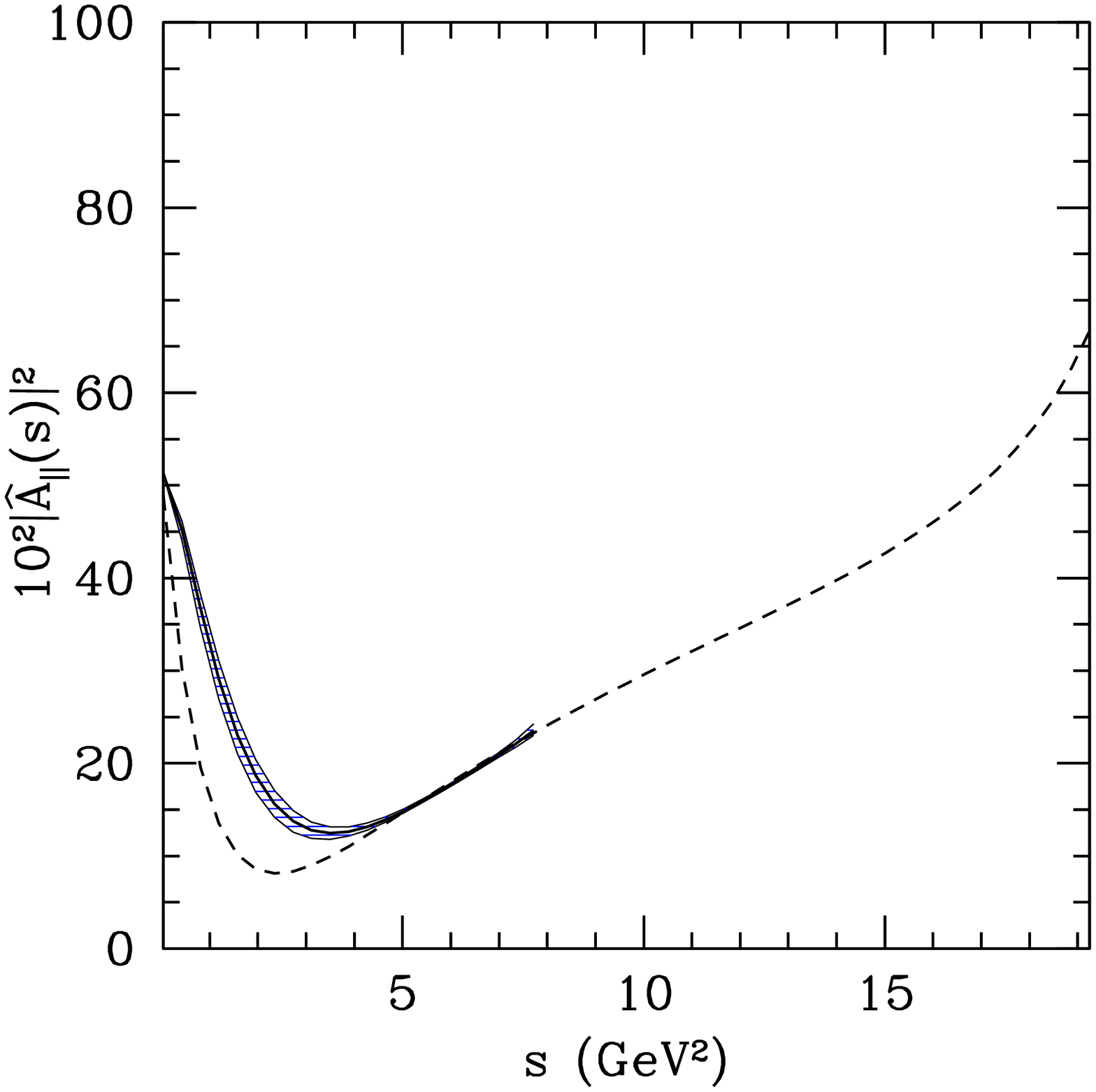}
\caption{ The helicity amplitude $|\hat {\cal A}_{\parallel}(s)|^2$
in $B \to K^* \ell^+ \ell^-$ at next-to-leading
order (solid center line) and leading order (dashed). The band for 
NLO reflects
the theoretical uncertainties from the input parameters.}
\label{figAII2}
\end{center}
\end{figure}
\begin{figure}[!hbtp]
\begin{center}
\includegraphics[width=12cm,height=9cm]{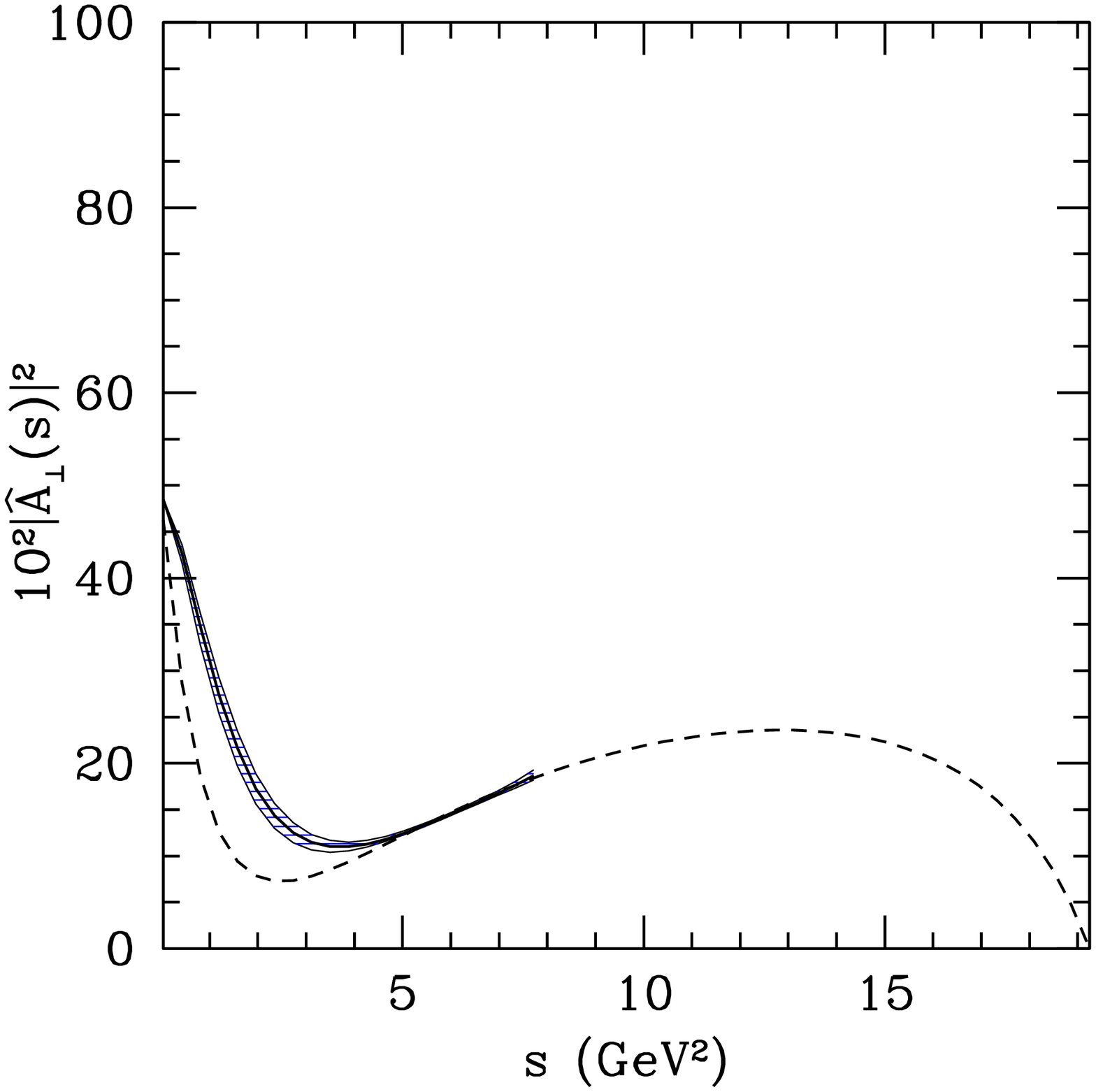}
\caption{ The helicity amplitude $|\hat {\cal A}_{\perp}(s)|^2$
in $B \to K^* \ell^+ \ell^-$ at next-to-leading order
(solid center line) and leading order (dashed). The band for NLO reflects 
theoretical uncertainties from input parameters.}
\label{figAperp2}
\end{center}
\end{figure}

Experimental results are conventionally expressed in terms of the spin
amplitudes $\hat {\cal A}_{0,\bot,\|}$ normalized to unity, with $|\hat
{\cal A}_0|^2+|\hat {\cal A}_\bot|^2+|\hat {\cal A}_\||^2=1$. We show the
polarization fractions, $\Gamma_0/\Gamma= |\hat {\cal A}_0(s)|^2$,
$\Gamma_{\parallel}/\Gamma= |\hat {\cal A}_{\parallel}(s)|^2$ and 
$\Gamma_{\perp}/\Gamma= |\hat {\cal A}_{\perp}(s)|^2$ in the
leading and next-to-leading order for the decay $B \to K^* \ell^+ \ell^-$
in Figs (\ref{figA02}), (\ref{figAII2}) and (\ref{figAperp2}),
respectively. Since the interference terms in the angular distribution are
limited to  \textrm{Re(${\cal A}_{||} {\cal A}_0^*$)},
\textrm{Im(${\cal A}_{\perp}
{\cal A}_0^*$)} and \textrm{Im(${\cal A}_{\perp} {\cal A}_{||}^*$)}, there
exists a phase ambiguity:
\begin{eqnarray}
\phi_{||} &\to& -\phi_{||}~,\\ 
\phi_{\perp} &\to& \pm \pi-\phi_{\perp}~,\\
\phi_{\perp}-\phi_{||} &\to& \pm \pi-(\phi_{\perp}-\phi_{||})~.
\label{eq:phidef}
\end{eqnarray}

\begin{figure}[H]
\begin{center}
\psfrag{s}{\hskip 0.cm $s\ (GeV^2)$}
\psfrag{bCfi2}{\hskip -0.1cm $\cos \phi_{||}(s)$}
\psfrag{bSfi2}{\hskip -0.2cm  $\sin \phi_{||}(s)$}
\epsfig{file=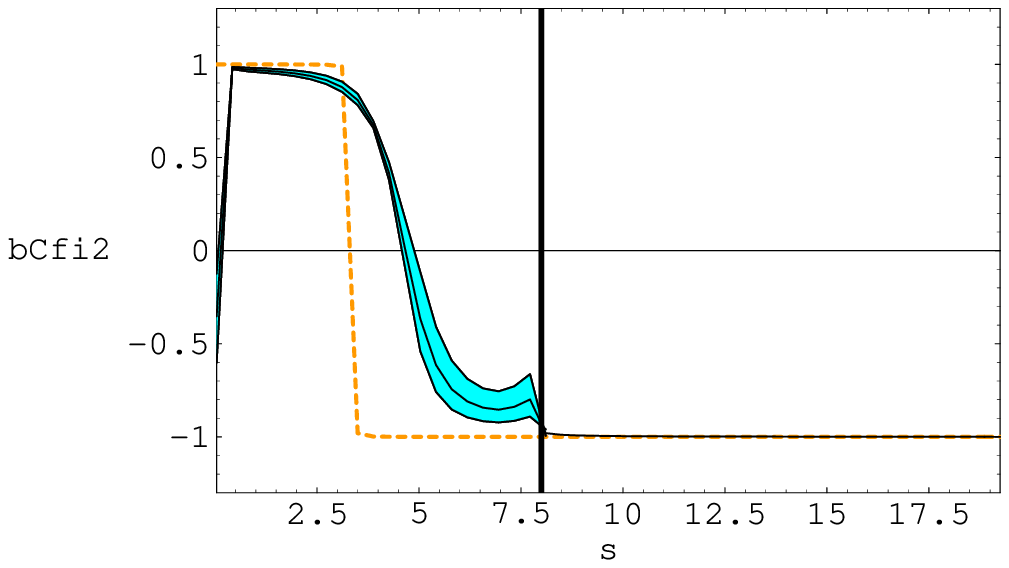,width=0.48\linewidth}
\hspace*{.2cm}
\epsfig{file=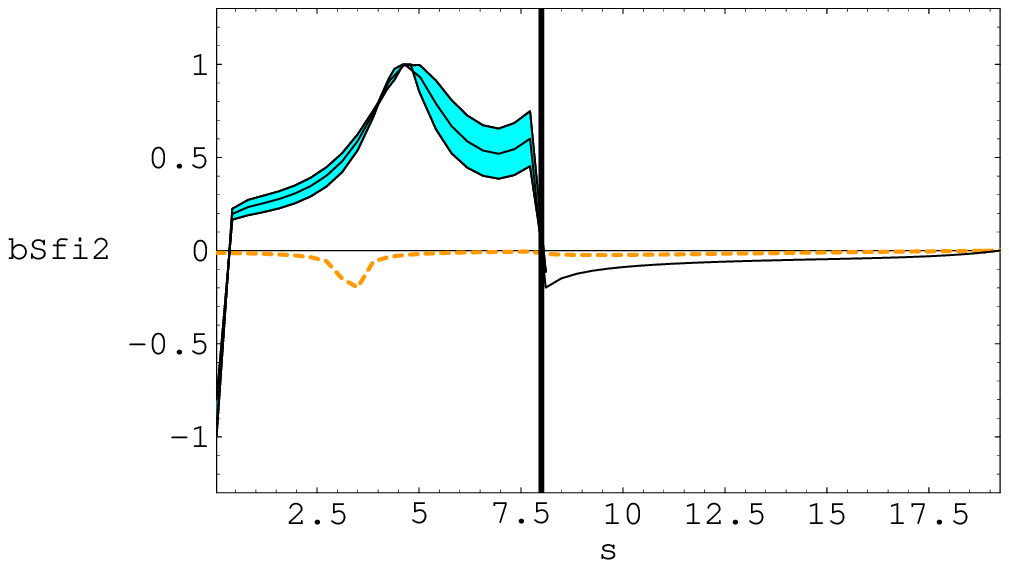,width=0.48\linewidth}
\caption{ 
The functions $\cos \phi_{||}(s)$ and $\sin \phi_{||}(s)$ at
next-to-leading order (solid center line) and leading order
(dashed). The band reflects all theoretical uncertainties from
parameters with most of the uncertainty due to the form factors
$\xi_{i}(0)$. The vertical line at \textrm{ s = 8 $GeV^2$} represents 
the domain of validity of the LEET approach in our case.}
\label{figphi-II}
\end{center}
\end{figure}

\begin{figure}[H]
\begin{center}
\psfrag{s}{\hskip 0.cm $s\ (GeV^2)$}
\psfrag{bCfiperp}{\hskip 0.cm $\cos \phi_{\perp}(s)$}
\psfrag{bSfiperp}{\hskip 0.2cm  $\sin \phi_{\perp}(s)$}
\epsfig{file=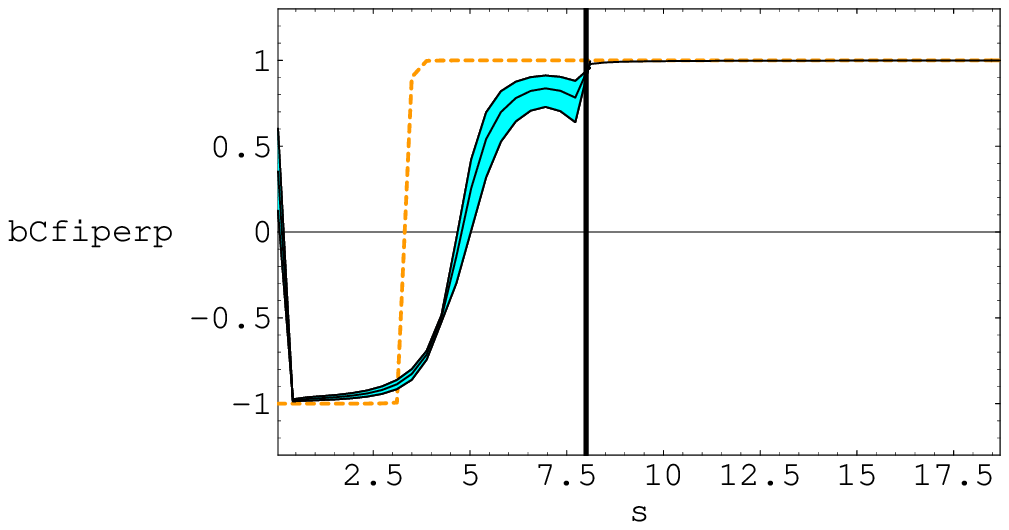,width=0.48\linewidth}
\hspace*{.2cm}
\epsfig{file=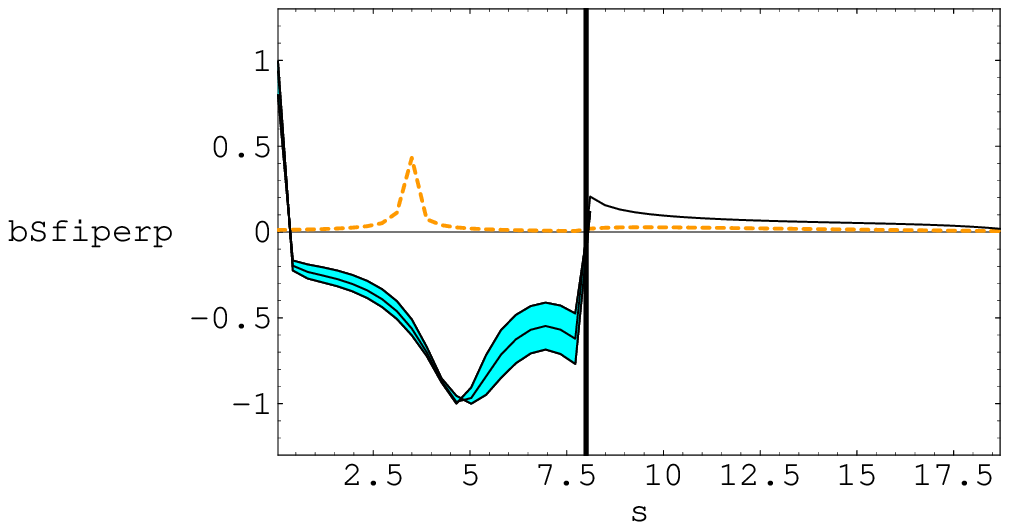,width=0.48\linewidth}
\caption{ 
The functions $\cos \phi_{\perp}(s)$ and $\sin \phi_{\perp}(s)$ at
next-to-leading order (solid center line) and leading order
(dashed). The band reflects all theoretical uncertainties from
parameters with most of the uncertainty due to the form factors
$\xi_{i}(0)$.The vertical line at \textrm{ s = 8 $GeV^2$} represents the
domain of validity of the LEET approach in our case.}
\label{figphi-perp}
\end{center}
\end{figure}

\noi To avoid this, we have plotted in Figs.~(\ref{figphi-II}) and
(\ref{figphi-perp}) the functions $\cos \phi_\parallel(s)$,
$\sin \phi_\parallel(s)$, and $\cos \phi_\perp(s)$, $\sin \phi_\perp(s)$,
respectively,  showing
their behaviour at the leading and next-to-leading order. 
The dashed lines in these figures correspond to using the LO amplitudes,
calculated in the LEET approach. In this order, the bulk of the parametric
uncertainty resulting from the form factors cancels. Although, strictly
speaking, the domain of validity
of the LEET-based distributions is limited by the requirement of large
energy of the $K^*$ (which we have translated into approximately $s < 8\
GeV^2$), we show this distribution for the entire $s$-region allowed
kinematically in $B \to K^* \ell^+ \ell^-$.
The shaded  curves correspond to using the NLO contributions in the LEET
approach. We compare the resulting amplitudes $\vert \hat {\cal A}_0
\vert^2$, $\vert
\hat {\cal A}_\perp \vert^2$, $\vert \hat {\cal A}_{\parallel} \vert^2$,
$\phi_{\parallel}(s)$, and $\phi_{\perp}(s)$ at the value $s=m_{J/\psi}^2$
with the corresponding results from the four experiments in
Table~\ref{others}. In comparing these results for the phases, we had to
make a choice between the two phase conventions shown in
Eq.~(\ref{eq:phidef}) and the phases shown in the last row of this table 
correspond to adopting the lower signs in these equations. 
We note that the short-distance amplitudes from the decay $B \to K^*
\ell^+ \ell^-$ are similar to their resonant counterparts measured in the
decay $B \to J/\psi K^*$. We also note that a helicity analysis of the
decay $B \to J/\psi K^*$ has been performed in the QCD factorization
approach by Cheng et al.~\cite{Cheng:2001ez}.  

\begin{figure}[H]
\psfrag{s}{\hskip 0.3cm $s\ (GeV^2)$}
\psfrag{b}{\hskip - 1.cm$\phi_{0}(s) $}
\begin{center}
\includegraphics[width=8cm,height=6cm]{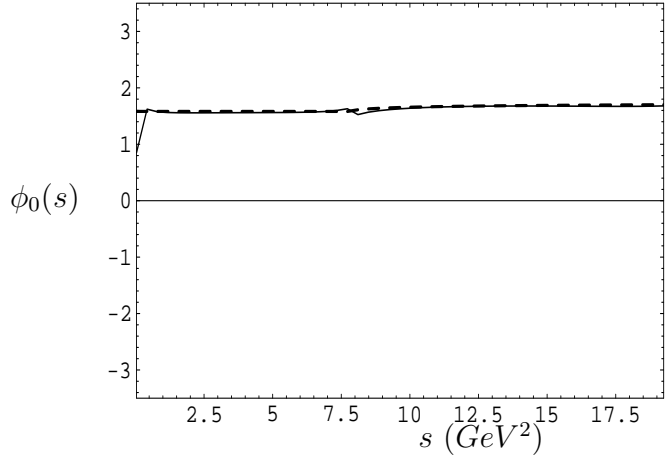} 
\caption{The phase $\phi_{0}(s) $ at next-to-leading
order (solid center line) and leading order (dashed).}
\label{figphi-0}
\end{center}
\end{figure}

The structures in the phases shown in  Fig.~(\ref{figphi-II}) and  
Fig.(\ref{figphi-perp}) deserve a closer look. We note that at
the leading order, the phases $\phi_{\perp}(s)$ and $\phi_{\parallel}(s)$
are given by the following expressions:
\begin{eqnarray}
\phi_{\perp}(s)&=& \textrm{Arg}\Big[{i\,\sqrt{\lambda}\over m_b\,
m_B\, \sqrt{s}}\Big\{s\, \cne+ 2\, m_b\, m_B\, \cse \Big\}\,
\xi_{\perp}(s)\Big] - \textrm{Arg}[{\cal A}_0(s)]~,\label{eq:phiperp}\\
\phi_{||}(s)&=& \textrm{Arg}\Big[{-i\,E_V\, \xi_{\perp}(s)\over m_b  
\,\sqrt{s}}\Big\{ \Big(s\, \cne+ 2\, m_b\, m_B\, \cse \Big)
- 2\, m_b\, m_B\, \Big(\cse +{s\over  2\, m_b\, m_B} Y(s)\Big)({m_V^2
\over m_B^2}) \Big\} \Big] \nn\\
&&- \textrm{Arg}[{\cal A}_0(s)],\label{eq:phiII}
\end{eqnarray}
\noi where we can neglect the term proportional to $(m_V^2/ m_B^2)$ in
the latter equation. The phase
$\phi_{0}(s)\equiv\textrm{Arg}[{\cal A}_0(s)]$ is
constant in the entire phase space, as shown in Fig.~(\ref{figphi-0}).
The functions in the square brackets in Eqs.~(\ref{eq:phiperp}) and
(\ref{eq:phiII}) are purely imaginary. However, due to the fact that in
the SM the coefficients $C_9^{\rm eff}$ and $C_7^{\rm eff}$ have opposite
signs, these phases become zero at a definite value of $s$, beyond which
they change sign, yielding a step-function behaviour, shown by
the dotted curves in the
functions $\cos \phi_{\parallel}(s)$ and $\cos \phi_{\perp}(s)$ in
Fig.~(\ref{figphi-II}) and Fig.~(\ref{figphi-perp}), respectively.
The position of the zero of the two functions, denoted, respectively,
 by $s^{\perp}_0$ and $s^{||}_0$, are given by solving the following
equations:
\begin{eqnarray}
 \textrm{Arg}\Big[{i\,\sqrt{\lambda}\over m_b\,
m_B\, \sqrt{s^{\perp}_0}}\Big\{s^{\perp}_0\, \cne(s^{\perp}_0)+ 2\,
m_b\, m_B\, \cse \Big\}\, \xi_{\perp}(s^{\perp}_0)\Big] &=& \phi_0(s^{\perp}_0)~,\label{eq:solphiperp}\\
 \textrm{Arg}\Big[{-i\,E_V\ \over m_b 
\,\sqrt{s^{||}_0}}\Big\{ s^{||}_0\, \cne(s^{||}_0)+ 2\, m_b\, m_B\, \cse \Big\}\, \xi_{\perp}(s^{||}_0) \Big] &=& \phi_0(s^{||}_0).\label{eq:solphiII}
\end{eqnarray}
\noi For the assumed values of the Wislon coefficients and other
parameters, the zeroes of the two functions, namely
$s^{||}_0$ and $s^{\perp}_0$, occur at around $s\simeq 3$ GeV$^2$, in the
lowest order, as can be seen in Figs. (\ref{figphi-II}) and
(\ref{figphi-perp}), respectively. The LO contributions in $\sin
\phi_{\parallel}(s)$ and $\sin \phi_{\perp}(s)$ are constant, with a value
around 0, with a small structure around $s\simeq 3\,
GeV^2$, reflecting the sign flip of the imaginary part in $
{\cal A}_{||}(s)$
(${\cal A}_{\perp}(s)$). At the NLO, the phases are influenced 
by the explicit $O(\alpha_s)$ contributions from the factorizable and
non-factorizable QCD corrections (see section 3), which also bring in
parametric uncertainties with them. The most important effect is that the
zeroes of the phases as shown for $\cos \phi_{\perp}(s)$ and
$\cos \phi_{\parallel}(s)$ are
shifted to the right, and the step-function type bahaviour of these
phases in the LO  gets a non-trivial shape.  Note
that in both figures a shoulder around $s\simeq 8\, GeV^2$ reflects charm
production whose threshold lies at $s = 4\, m_c^2$.

\begin{table}
\begin{center}
\begin{tabular}{|c|c|c|c|c|c|}   
\hline
Group & $|\hat {\cal A}_0|^2$ & $|\hat {\cal A}_{\perp}|^2$ & $|\hat
{\cal A}_{||}|^2$ & 
$\phi_{\perp}$ & $\phi_{\parallel}$
\\ \hline
CLEO\cite{ref:cleo} &
$0.52\pm0.08$ & $0.16\pm0.09$ & $0.32\pm0.12$ &
$-3.03\pm0.46$ & $-3.00\pm 0.37$ \\
CDF\cite{ref:cdf} &    
$0.59\pm0.06$ & $0.13^{+0.13}_{-0.11}$ &$0.28\pm0.12$ &
$-2.58\pm0.54$ & $-2.20\pm 0.47$ \\
BaBar\cite{ref:babar} &
$0.60\pm0.04$ & $0.16\pm0.03$ & $0.24\pm0.04 $& 
$-2.97\pm0.17$ &$-2.50\pm 0.22$ \\
Belle \cite{ref:belle}&
$0.60\pm0.05$ & $0.19\pm0.06$ & $0.21\pm0.08 $&
$-3.15\pm0.21$ & $-2.86\pm0.25$ \\
This Work &$0.51$ &$0.21$ & $0.28$ &$-3.25$ &$-3.04$ \\
\hline
\end{tabular}
\caption{Current measurements of the decay amplitudes in the transversity
basis for the decay $B\rightarrow J/\psi   K^*$ .
The corresponding amplitudes for the
non-resonant decay $B \to K^* \ell^+ \ell^-$ worked out
in this paper in the LO approximation at $m_{\ell^+
\ell^-}^2=m_{J/\psi}^2$ are given in the last row. \label{others}}
\end{center} \end{table}
%
%
%
\section{Decay Distributions in $B \rightarrow \rho \ell \bar{\nu_{\ell}}$}
\hspace*{\parindent}
\noi The differential decay rate for $B\rightarrow \rho (\rightarrow
\pi^{+} \pi^{-}) \ell \bar{\nu_{\ell}}$ can be  expressed as follows
\cite{Korner:1989qb,Korner:1989tb,Richman:wm}:
\begin{eqnarray}
\label{4diff}
{d^4\Gamma  \over ds\ d\cos\theta_{\rho}\ d\cos\theta_{+}\ d\phi} &=& {3
\over 8 (4\pi)^4} G_{F}^2 |V_{ub}|^2 {\sqrt{\lambda} s \over  m_{B}^3} {\it{{\cal B}(\rho \rightarrow \pi^+ \pi^-)}}\nn \\
&& \Big \{(1 - \ \cos\theta_{+})^2 \sin^2\theta_{\rho}\ |H_{+}(s)|^2\nn \\  
&& + (1 +\ \cos\theta_{+})^2  \sin^2\theta_{\rho}\  |H_{-}(s)|^2\nn \\
&& + 4\ \sin^2\theta_{+}\  \cos^2\theta_{\rho}\  |H_{0}(s)|^2\nn  \\
&& - 4\ \  \sin\theta_{+}\ (1 -  \cos\theta_{+})\ \sin\theta_{\rho}\ \cos\theta_{\rho}\ \cos\phi\  H_{+}(s)  H_{0}(s)\nn \\
&& + 4\ \  \sin\theta_{+}\ (1 + \cos\theta_{+})\  \sin\theta_{\rho}\ \cos\theta_{\rho}\ \cos\phi\ H_{-}(s) H_{0}(s)\nn \\
&& -2\ \sin^2\theta_{+}\ \sin^2\theta_{\rho}\ \cos2\phi\ H_{+}(s)
H_{-}(s)\Big \}~. \label{diffrho}
\end{eqnarray}
The three angles $\theta_+$, $\theta_{\rho}$ and $\phi$  are defined
as follows: $\theta_+$ is defined by the direction between the charged lepton
and the recoiling vector meson measured in the $W$ rest frame, the polar
angle $\theta_{\rho}$ is defined by the directions of the
$\pi^{+}$ (or $\pi^{-}$) and the vector meson in the
parent meson's rest frame, and the azimuthal angle $\phi$ is the angle 
between the two planes, defined by the momenta of $\pi^{+} \pi^{-}$ and the
lepton pair $\ell \bar{\nu}$.

\begin{figure}
\psfrag{c}{\hskip 0.3cm $s\ (GeV^2)$}
\psfrag{a}{\hskip -4.2cm $d^2{\cal B}_{|H_{-}|^2}/ ds\ d\cos\theta_{+}
|V_{ub}|^2$}
\psfrag{b}{$ \cos\theta_{+}$}
\begin{center}
\includegraphics[width=12cm,height=9cm]{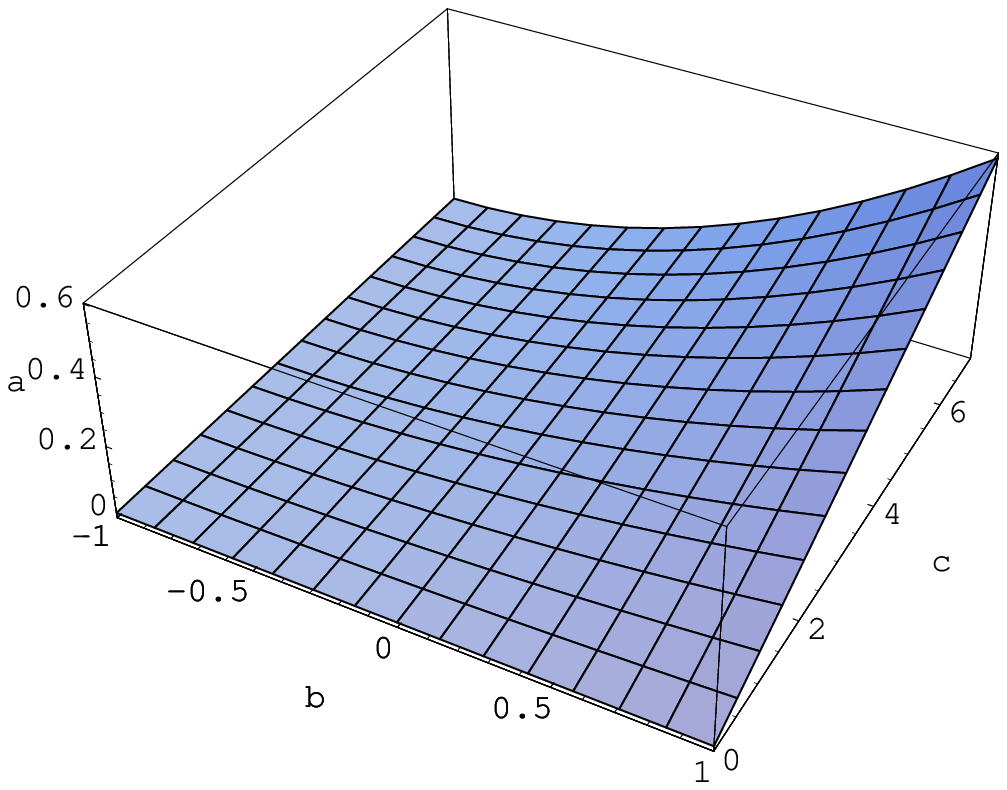}
\caption{Partial Dalitz distribution ${d^2{\cal B}_{|H_{-}|^2}(B \rightarrow 
 \rho \ell \bar{\nu_\ell}) \over
d\cos\theta_{+}\ ds}$.}
\label{dalitzdsdcosthetaH-}
\psfrag{c}{\hskip 0.3cm $s\ (GeV^2)$}
\psfrag{a}{\hskip -3.5cm $d^2{\cal B}_{|H_{0}|^2}/ ds\ d\cos\theta_{+}
|V_{ub}|^2$}
\psfrag{b}{$ \cos\theta_{+}$}
\includegraphics[width=12cm,height=9cm]{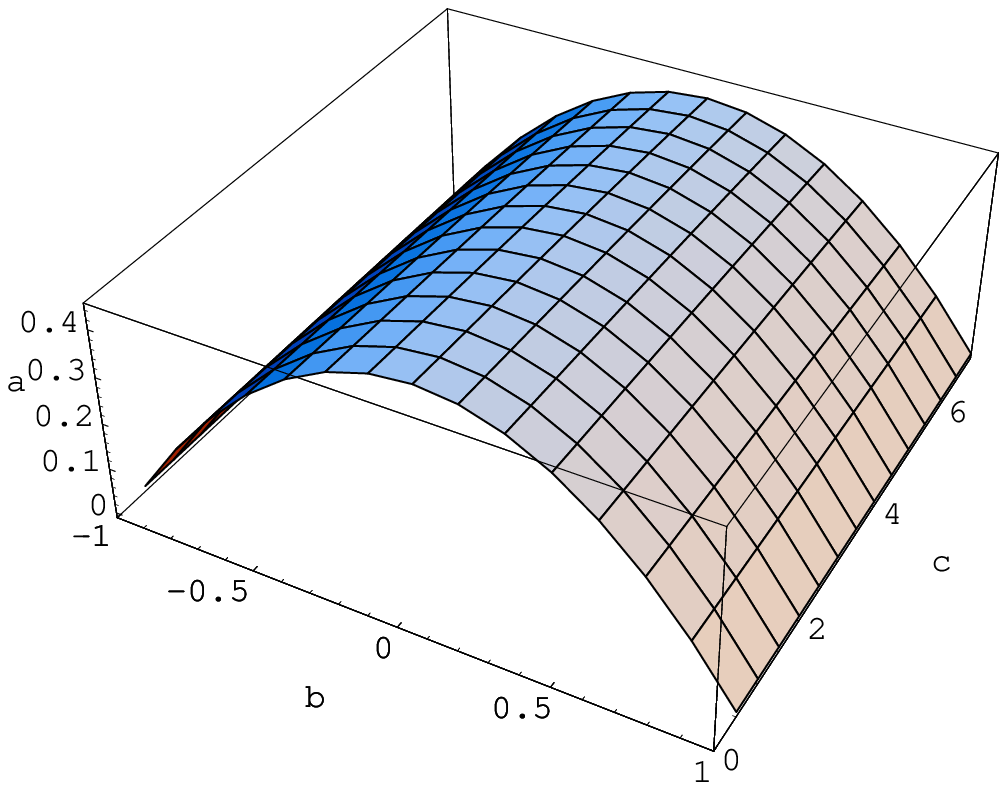}
\caption{Partial Dalitz distribution ${d^2{\cal B}_{|H_{0}|^2}(B
\rightarrow  \rho \ell \bar{\nu_\ell}) \over d\cos\theta_{+}\ ds}$.}
\label{dalitzdsdcosthetaH0}
\end{center}
\end{figure}

\begin{figure}
\psfrag{c}{\hskip 0.3cm $s\ (GeV^2)$}
\psfrag{a}{\hskip -3.5cm $d^2{\cal B}/ ds\ d\cos\theta_{+} |V_{ub}|^2$}
\psfrag{b}{$ \cos\theta_{+}$}
\begin{center}
\includegraphics[width=12cm,height=9cm]{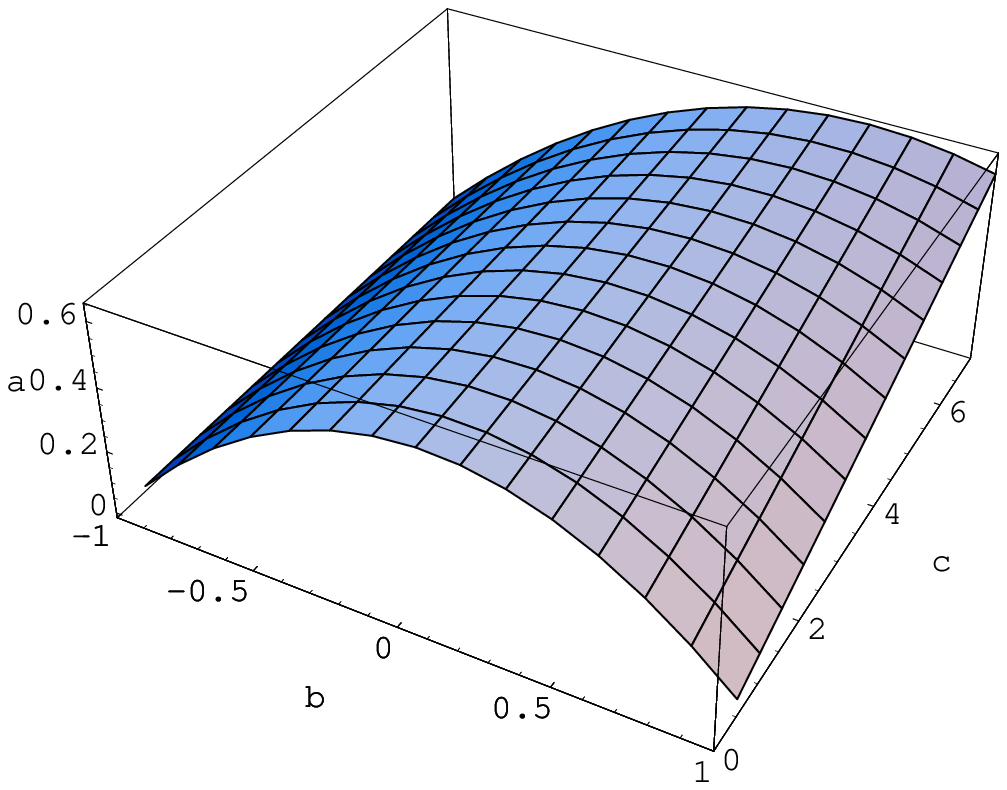}
\caption{Dalitz distribution ${d^2{\cal B}(B \rightarrow  \rho \ell \bar{\nu_\ell}) \over \d\cos\theta_{+}\ ds}$ .}
\label{dalitzdsdcostheta+}
\end{center}
\end{figure}

\begin{figure}
\psfrag{c}{\hskip 0.3cm $s\ (GeV^2)$}
\psfrag{a}{\hskip -3.5cm $d^2{\cal B}/ ds\ d\cos\theta_{\rho} |V_{ub}|^2$}
\psfrag{b}{$ \cos\theta_{\rho}$}
\begin{center}
\includegraphics[width=12cm,height=9cm]{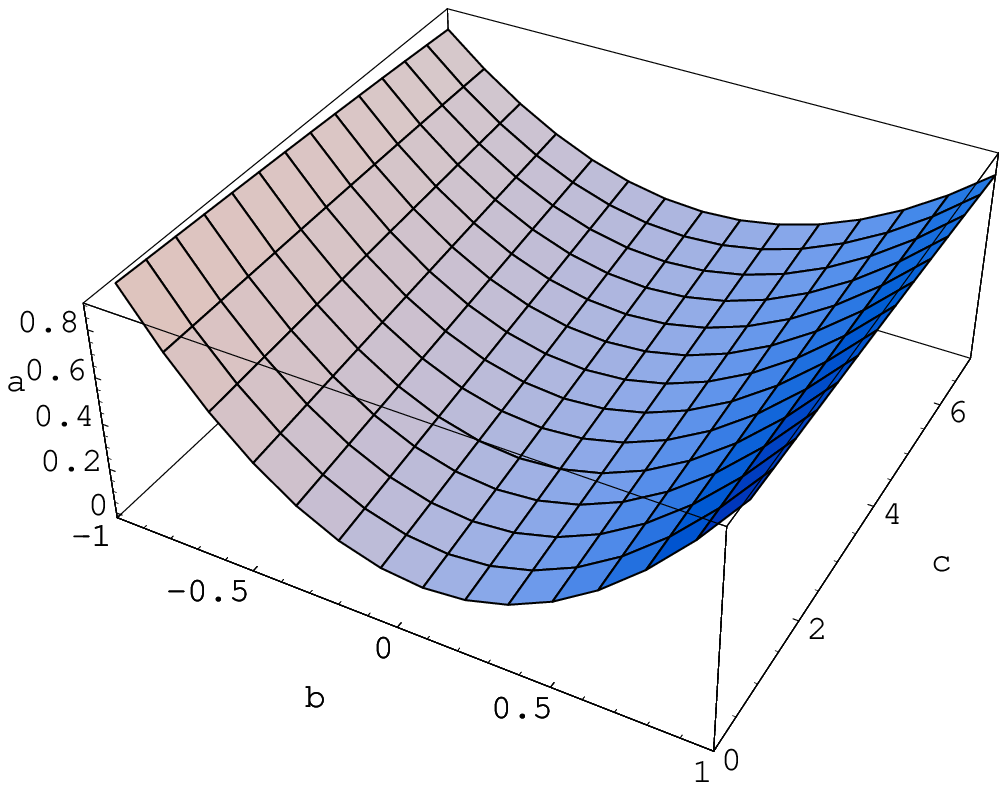}
\caption{Dalitz distribution  ${d^2{\cal B}(B \rightarrow  \rho \ell \bar{\nu_\ell}) \over \cos\theta_{\rho}\ ds}$.}
\label{dalitzdsdcosthetav}
\end{center}
\end{figure}

The helicity  amplitudes can in turn be related to the two
axial-vector form factors, $A_{1}(s)$ and $A_{2}(s)$, and the vector
form factor, $V(s)$, which appear in the hadronic current 
\cite{Richman:wm}:
\begin{eqnarray}
H_{\pm}(s) &=& (m_{B} + m_{\rho})\ A_{1}(s) \mp 2\ { \sqrt{\lambda}
\over m_{B} + m_{\rho}}\ V(s)~, \label{Hpm} \\
H_{0}(s) &=& {1 \over 2m_{\rho} \sqrt{s}}\ \Big[(m_{B}^2 - m_{\rho}^2-
s)(m_{B} + m_{\rho})\ A_{1}(s) - 4\ {\lambda \over m_{B} + m_{\rho}}\
 A_{2}(s) \Big] \label{H0}~.
\end{eqnarray}
Using Eqs.(\ref{A1}),(\ref{A2}) and (\ref{V}) in Eqs.~(\ref{H0}) and
(\ref{Hpm}), we obtain the helicity  amplitudes in the large energy Limit:
\begin{eqnarray}
H_{\pm}(s) &=& 2 \left[ E_{\rho} \mp {\sqrt{\lambda} \over m_B} \right ] \
 \xi_{\perp}(s)~,\label{HsemiLETpm}\\
H_{0}(s) &=& {1 \over m_{B}\ m_{\rho}\ \sqrt{s}}  \Big[ m_B\ E_{\rho}\
 (m_{B}^2
 - m_{\rho}^2 - s)  - 2\ \lambda  \Big]\ \xi_{\perp}(s) \label{HsemiLET0} \\
&& + {2\ \lambda \over m_B E_{\rho} \sqrt{s}}\ \xi_{||}(s)~.\nn 
\end{eqnarray}
We give below the double differential decay rate (Dalitz distribution) for
$B\rightarrow \rho(\rightarrow \pi^{+} \pi^{-}) \ell \bar{\nu}$ in the
variables $(s,\phi)$, $(s,\cos \theta_{+})$ and $(s,\cos \theta_\rho)$,
giving also the expressions for the individual contributions from the
Helicity-$0$ and Helicity-$-1$ amplitudes for the Dalitz distribution in
$(s,\cos \theta_{+})$:
 \begin{eqnarray}
{d^2{\cal B} \over d\phi\ ds} &=&\tau_B {G_F^2\ s\  \sqrt{\lambda}
 \over 192 m_{B}^3 \pi^4} |V_{ub}|^2 ({\it{{\cal B}(\rho \rightarrow
 \pi^+ \pi^-)}})~,
\label{dgamadsdchi} \\ 
&& \Big \{  |H_0(s)|^2 +  |H_-(s)|^2 + |H_+(s)|^2 -   \cos 2\phi \ H_-(s)\
  H_+(s)\Big\}~,\nn \\
{d^2{\cal B}_{|H_{-}|^2} \over d\cos\theta_{+}\ ds} &=&  \tau_B {G_{F}^2\ s\
\sqrt{\lambda} \over 256 m_{B}^3 \pi^3} |V_{ub}|^2
 ({\it{{\cal B}(\rho \rightarrow \pi^+ \pi^-)}})\Big
 \{(1 + \cos\theta_{+})^2\ |H_{-}(s)|^2 \Big\}~,\nn \\
{d^2{\cal B}_{|H_{0}|^2} \over d\cos\theta_{+}\ ds} &=&  \tau_B {G_{F}^2\ s\
\sqrt{\lambda} \over 256 m_{B}^3 \pi^3} |V_{ub}|^2 ({\it{{\cal B}(\rho \rightarrow
 \pi^+ \pi^-)}}) \Big \{2\  \sin^2\theta_{+}\  |H_0(s)|^2\Big\}~,\nn \\
{d^2{\cal B} \over d\cos\theta_{+}\ ds} &=& \tau_B {G_{F}^2\ s\ \sqrt{\lambda}
\over 256 m_{B}^3 \pi^3} |V_{ub}|^2 ({\it{{\cal B}(\rho \rightarrow \pi^+ \pi^-)}})
~,\nn \\ 
&& \Big \{2\  \sin^2\theta_{+}\  |H_0(s)|^2+(1 -  \ \cos\theta_{+})^2\
 |H_{+}(s)|^2 + (1 + \cos\theta_{+})^2\ |H_{-}(s)|^2 \Big\}\nn \\
{} & = &{d^2{\cal B}_{|H_{0}|^2} \over d\cos\theta_{+}\ ds}+{d^2{\cal B}_{|H_{+}|^2} \over
d\cos\theta_{+}\ ds}+{d^2{\cal B}_{|H_{-}|^2} \over d\cos\theta_{+}\
ds}~,\label{dgamadsdcos+}\\
{d^2{\cal B} \over d\cos\theta_{\rho}\ ds} &=&  \tau_B {G_{F}^2\ s\
\sqrt{\lambda}\over 128 m_{B}^3 \pi^3} |V_{ub}|^2 ({\it{{\cal B}(\rho \rightarrow
 \pi^+ \pi^-)}})\label{dgamadsdcosthetav} \\ 
&&  \Big \{ 2\ \cos^2\theta_{\rho}\ |H_0(s)|^2 +\   \sin^2\theta_{\rho}\
 \Big( |H_{+}(s)|^2 + |H_{-}(s)|^2\Big) \Big\}~.\nn 
\end{eqnarray}
In Figs.~(\ref{dalitzdsdcosthetaH-}), (\ref{dalitzdsdcosthetaH0}),
(\ref{dalitzdsdcostheta+})
and (\ref{dalitzdsdcosthetav}), we show, respectively,
the Dalitz distributions ($d^2{\cal B}_{|H_{-}|^2}/ d\cos\theta_{+}\ ds$),
($d^2{\cal B}_{|H_{0}|^2}/ d\cos\theta_{+}\ ds$), ($d^2{\cal B}/
d\cos\theta_{+}\ ds$) and ($d^2{\cal B}/ d\cos\theta_{\rho}\ ds$).

\begin{figure}
\psfrag{a}{$s\ (GeV^2)$}
\psfrag{b}{\hskip -3.cm $d{\cal B}_{|H_{-}|^2}/ ds\ |V_{ub}|^2 $}
\psfrag{c}{}
\begin{center}
\includegraphics[width=12cm,height=9cm]{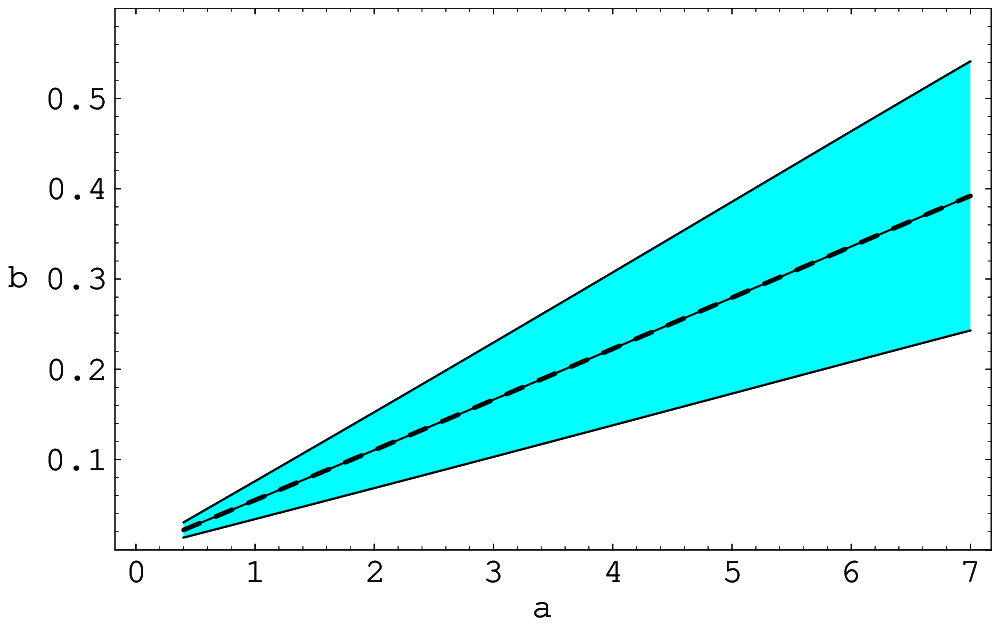}
\caption{The dilepton invariant mass distribution
$d{\cal B}_{|H_{-}|^2}/ ds$ for $ B \rightarrow \rho \ell \bar{\nu}$ 
at next-to-leading order (solid center line) and leading order
(dashed). The band reflects  theoretical uncertainties from input
parameters.}
\label{dBrH-1rho}
\end{center}
\end{figure}

\begin{figure}
\psfrag{a}{\hskip -1.cm $s\ (GeV^2)$}
\psfrag{b}{\hskip -3.cm $d{\cal B}_{|H_{0}|^2}/ ds\ |V_{ub}|^2 $}
\begin{center}
\includegraphics[width=12cm,height=9cm]{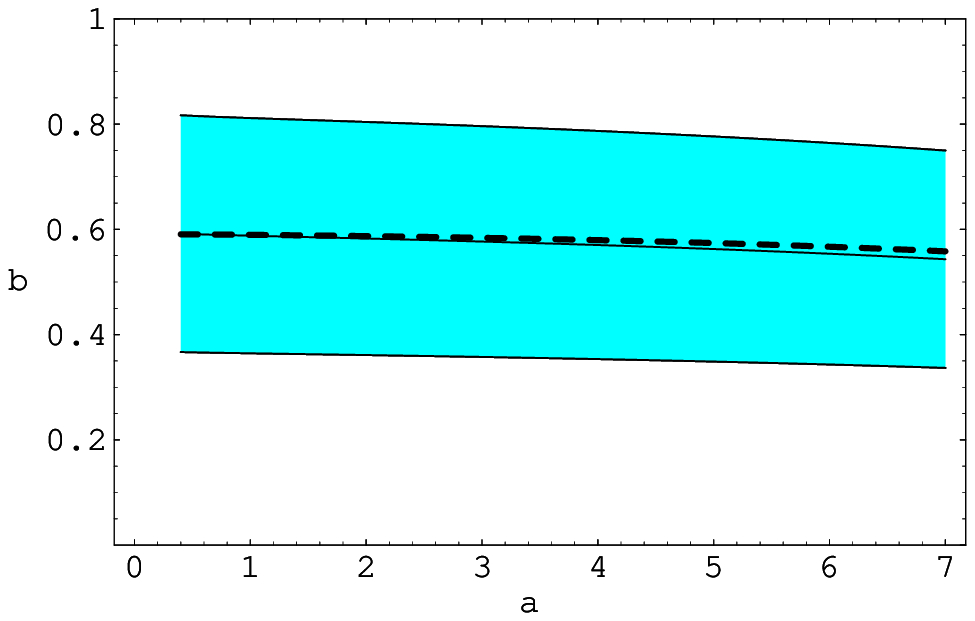}
\caption{The dilepton invariant mass distribution
$d{\cal B}_{|H_{0}|^2}/ ds$ for $ B \rightarrow \rho \ell \bar{\nu}$ at
 next-to-leading order (solid center line) and leading order (dashed). 
The band reflects theoretical uncertainties from input
parameters.}
\label{dBrH0rho}
\end{center}
\end{figure}

\begin{figure}
\psfrag{a}{$s\ (GeV^2)$}
\psfrag{b}{\hskip -2.5cm $d{\cal B}/ ds\ |V_{ub}|^2 $}
\psfrag{c}{}
\begin{center}
\includegraphics[width=12cm,height=9cm]{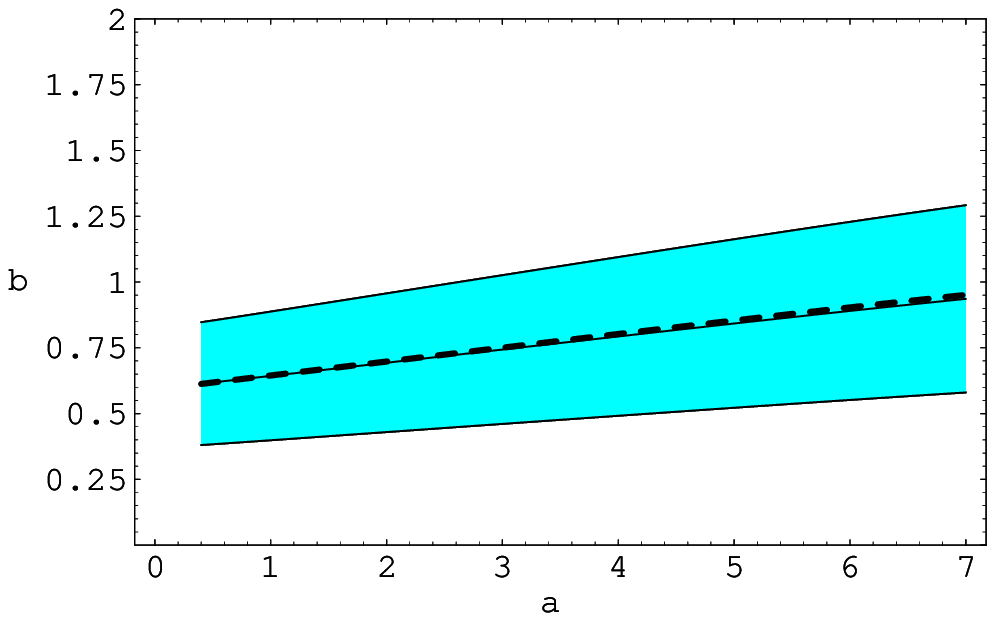}
\caption{The dilepton invariant mass distribution for $ B \rightarrow
\rho \ell \bar{\nu}$
at next-to-leading order (solid center line) and leading
order (dashed). The band reflects  theoretical uncertainties from
input parameters.}
\label{dBrrho}
\end{center}
\end{figure}

\noi Integrating out the angle $\theta_+$, $\theta_{\rho}$ and $\phi$ from
Eq.~(\ref{diffrho}), we obtain  the total branching decay
rate: 

\begin{eqnarray}
{d{\cal B} \over ds} &=& \tau_B {G_F^2\ s\ \sqrt{\lambda} \over 96 m_{B}^3 \pi^4} |
V_{ub}|^2 ({\it{{\cal B}(\rho \rightarrow \pi^+ \pi^-)}}) \Big\{|H_0(s)|^2 +
 |H_{+}(s)|^2 + |H_{-}(s)|^2 \Big\} \label{dgamads}\\ 
{} &=& {d{\cal B}_{|H_{0}|^2} \over ds}+{d{\cal B}_{|H_{+}|^2} \over ds}+{d{\cal B}_{|H_{-}|^2}
 \over ds} ~.\nn
\end{eqnarray}
\noi Just as in the decay $B \to K^* \ell^+ \ell^-$, the contribution from
the $\vert H_{+}(s)\vert^2$ is negligible, and we do not show it here.
The contributions from the $\vert H_{-}(s)\vert^2$, $\vert
H_{0}(s)\vert^2$ and the total are shown in Figs.~(\ref{dBrH-1rho}),
(\ref{dBrH0rho}) and (\ref{dBrrho}), respectively. The impact of the NLO
correction on the various branching ratios in $B \rightarrow \rho \ell
\nu_\ell$ is less significant than in the $B \rightarrow K^* \ell^+
\ell^- $ decay,  reflecting the absence of the penguin-based amplitudes
in the former decay.

Concerning the $B \rightarrow \rho \ell \nu_\ell$ form factors,
one has to consider the SU(3)-breaking effects in relating them to the
corresponding form factors in $B \to K^* \ell^+ \ell^-$. For the form
factors in full QCD, they have been evaluated within the Light-cone QCD
 sum-rule approach~\cite{Ali:vd,Ball:1998kk,Ball:2004rg},
 Lattice-QCD~\cite{DelDebbio:1997kr}, and in quark models of the more
recent vintage~\cite{Beyer:1998ka,Gilani:2003hf}.
 Based on the Light-cone  QCD sum-rule approach, we estimate the
 SU(3)-breaking in the ratio of the LEET-based form factors at $s=0$
as
\begin{equation}
\zeta_{SU(3)}={\xi^{(K^*)}_{\perp,||}(0)\over
\xi^{(\rho)}_{\perp,||}(0)}= 1.3 \pm 0.06~.
\end{equation}
With this and $\xi^{(K^*)}_{\perp}(0)$ from Table \ref{parameters},
we obtain:
\begin{equation}
\xi^{(\rho)}_\perp(0)=0.22 \pm 0.04~. 
\end{equation}
The results for the various helicity amplitudes $H_{+}(0)$, $H_{-}(0)$,
$\sqrt{s} H_{0}(0)$, and the ratio $H_{-}(0)/H_{+}(0)$ in the decay
$B \to \rho \ell \nu_\ell$  are given in Table~\ref{Hats0},
and compared with some other estimates of the same in the literature. We recall that
in our approach both $\xi^{(K^*)}_{\perp}(0)$ and $\xi^{(\rho)}_\perp(0)$
are smaller due to the normalization of the former from data on $B \to
K^*\gamma$ decay rate. This is reflected  in the smaller values of the
helicity amplitudes in this work compared to the other approaches.

To get the $B \rightarrow  \rho\ \ell\ \nu_\ell$ form factors at
$s \neq 0$, we use the same parametrization as the one for the
$B \rightarrow K^* \ell^+ \ell^-$ form factors:
\begin{equation}
\xi^{(\rho)}_{\perp,||}(s)={\xi^{(K^*)}_{\perp,||}(s) \over
\zeta_{SU(3)}}~.
 \label{eq:paraball}
\end{equation}
This parametrization has been used in calculating the differential branching
 ratios shown for the restricted region $s \leq 7$ GeV$^2$ in 
 Figs.~(\ref{dBrH-1rho}), (\ref{dBrH0rho}) and (\ref{dBrrho}).
 While we
do not insist on using our approach to cover the small
 $E_\rho$-region (and, hence higher values of $s$) in the decay $B \to \rho \ell \nu_\ell$,
but for the sake of comparison with recent data and 
 some existing estimates of the various helicity amplitudes
in the literature, we use the $s$-dependence in (\ref{eq:paraball})
to estimate $H_i(s)$ in the entire kinematic region.
 Numerical results from our work are shown in the first row
 of Table~\ref{Hatsmax} for $s=s_{\rm max}$.
The equality $H_{+}(s_{\rm max})=H_{-}(s_{\rm max})=H_{0}(s_{\rm max})$
is imposed by the kinematics of the decay $B \to \rho \ell \nu_\ell$, but we
 note that our results are numerically smaller than the ones following from
 other form factor models 
shown in this table.  
 This is again to be traced back to our normalization of the function
$\xi_{\perp}^{K^*}(0)$ and (\ref{eq:paraball}).
 In Table~\ref{gammas}, 
we present various reduced partial widths for $B \to \rho \ell \nu_\ell$
for the longitudinal part $\Gamma_L=\Gamma_0$, transverse part
 $\Gamma_T=\Gamma_{-}+\Gamma_{+}$,
the total reduced width $\Gamma_{\rm Total}$,
factoring out the CKM matrix element $\vert V_{ub}\vert^2$,
 and the ratio
 $\Gamma_L/\Gamma_T$, and compare them with the corresponding estimates
in the literature. Our values for the total reduced decay widths are
smaller than the other five shown in Table ~\ref{gammas}.
For the form factor model in Ref.~\cite{Gilani:2003hf}, we show a detailed
comparison and note that 
 our estimate of $\Gamma_T$ (derived from LEET and data) is significantly
 smaller than 
in this model,
but our $\Gamma_L$, with input from the Light-cone QCD sum rules, is
comparable to
the one in  Ref.~\cite{Gilani:2003hf}. 
 This has the
consequence that in our approach $\Gamma_L $ and  $\Gamma_T$ are about
equal, as opposed to theirs where $\Gamma_L $ is significantly smaller than
$\Gamma_T$, which will be tested in future experiments.

 Finally,
we note that the smaller values of the form factors in the LEET-based approach
used in this work lead to estimates of $\vert V_{ub} \vert$ from the 
measured exclusive
decay rate for $B \to \rho \ell \nu_\ell$~\cite{Aubert:2005cd}, which are
in agreement with the PDG 2004 
 average~\cite{Eidelman:2004wy} $\vert V_{ub} \vert=(3.67 \pm 0.47) \times
10^{-3}$. This can also be seen as follows:
Using the central values of $\vert V_{ub}\vert$
and the lifetime $\tau_{B^0}=1.536$~ps from  PDG 2004,
and the reduced width in the LEET approach $\Gamma_{\rm Total}=10.49$ ps$^{-1}$ from
Table~\ref{gammas},  we get
 ${\cal B}(B \to \rho \ell \nu_\ell)=2.18\times 10^{-4}$,
in comfortable agreement with the measured branching ratio by the BABAR
 collaboration~\cite{Aubert:2005cd}
 ${\cal B}(B \to \rho \ell \nu_\ell)=(2.14 \pm 0.21\pm 0.48)\times 10^{-4}$.
In contrast, the allowed range of $\vert V_{ub}\vert$ 
 obtained recently by the BABAR collaboration~\cite{Aubert:2005cd} using
 the form factors from the Light-cone QCD sum-rules~\cite{Ball:2004rg}
is  $\vert V_{ub} \vert=(2.82 \pm 0.18 \pm 0.30 \pm 0.18) \times
10^{-3}$, with
similar values for the quark-model based form factors~\cite{ISGW2}.
This is significantly smaller than the PDG 2004
 average~\cite{Eidelman:2004wy} $\vert V_{ub} \vert=(3.67 \pm 0.47) \times
10^{-3}$. 
Thus, it is evident that, unless the corrections to the LEET-based approach
are very significant, data for the radiative decays
 $B \to (K^*,\rho,\omega) \gamma$, but also for the semileptonic decay
$B \to \rho \ell \nu_\ell$, favour smaller form factors than
obtained in the QCD sum rule approaches or the quark models.
%
\begin{table}[H]
\begin{center}
\begin{tabular}{|c|c|c|c|c|}
\hline
& $H_{+}\left( 0\right) $ & $H_{-}\left( 0\right) $ & $H_{-}\left( 0\right)
/H_{+}\left( 0\right) $ & $\sqrt{s}H_0\left( 0\right) $ \\ \hline
This Work& 0.048 & 2.274 & 47.002 & 7.459 \\ \hline
Ref.\cite{Gilani:2003hf} & 0.256 & 3.253 & 12.714 & 8.737 \\ \hline
QM \cite{Beyer:1998ka} & 0.163 & 2.942 & 18.014 & 8.202 \\ \hline
LCSR \cite{Ball:1998kk} & 0.055 & 3.103 & 56.767 & 10.158 \\ \hline
Lattice \cite{DelDebbio:1997kr} & 0.0898 & 4.224 & 47.020 & 10.914 \\ \hline
\end{tabular}
\caption{Various helicity amplitudes for $B \to \rho \ell \nu_\ell$  
calculated at $s=0$, using different models.}
\label{Hats0}
\end{center}
\end{table}
\begin{table}[H]
\begin{center}
\begin{tabular}{|c|c|c|c|c|}
\hline
& $H_{+}\left( s_{\max }\right) $ & $H_{-}\left( s_{\max }\right) $ & $%
H_0\left( s_{\max }\right) $ & $\sqrt{s}H_0\left( s_{\max }\right) $
\\ \hline
This Work& 1.919 & 1.919 & 1.919 & 8.656 \\ \hline
Ref.\cite{Gilani:2003hf} & 2.362 & 2.362 & 2.362 & 10.652 \\ \hline
QM \cite{Beyer:1998ka} & 2.952 & 2.952 & 2.952 & 13.313 \\ \hline
LCSR \cite{Ball:1998kk} & 2.782 & 2.782 & 2.782 & 12.549 \\ \hline
Lattice \cite{DelDebbio:1997kr} & 2.826 & 2.826 & 2.826 & 12.744 \\ \hline
\end{tabular}
\caption{Various helicity amplitudes for $B \to \rho \ell \nu_\ell$  
calculated at $s=s_{\max}$, using different models.}
\label{Hatsmax}
\end{center}
\end{table}
\begin{table}[H]
\begin{center}
\begin{tabular}{|c|c|c|c|c|c|c|}
\hline
& $\Gamma _{-}$ & $\Gamma _{+}$ & $\Gamma _L=\Gamma _0$ & $\Gamma _{T}=
\Gamma _{-}+\Gamma _{+}$ & $%
\Gamma _{\mathrm{Total}}$ & $\Gamma _L/\Gamma _T$ \\ \hline
This Work&4.978& 0.084& 5.423& 5.062& 10.49&1.07\\ \hline
Ref.\cite{Gilani:2003hf} & 9.662 & 0.059 & 6.095 & 9.721 & 15.82 & 0.63 \\ 
\hline
QM \cite{Beyer:1998ka} & - & - & - & - & 15.8$\pm $2.3 & 0.88$\pm 0.08$ \\ 
\hline
LCSR \cite{Ball:1998kk} & - & - & - & - & 13.5$\pm 4.0$ & 0.52$\pm $0.08 \\ 
\hline
Lattice \cite{DelDebbio:1997kr} & - & - & - & - & 16.5$_{-2.3}^{+3.5}$ & 0.80$%
_{-0.03}^{+0.04}$ \\ \hline
ISGW2 \cite{ISGW2} & - & - & - & - & 14.2 & 0.3 \\ \hline
\end{tabular}
\caption{Reduced partial and total decay rates  for $B \to \rho \ell \nu_\ell$
in units of ps$^{-1}$, the ratio $\Gamma_L/\Gamma_T$, obtained in our
 approach, and using different models for form factors. To get the decay widths,
one has to multiply the entries in this table with $\vert V_{ub}\vert^2$.}
\label{gammas}
\end{center}
\end{table}

\section{Determination of $|V_{ub}|/|V_{ts}|$ from $B \to \rho \ell
\nu_\ell$ and\\ $B \to K^* \ell^+ \ell^-$ Decays}
\begin{figure} 
\psfrag{a}{ $s\ (GeV^2)$}
\psfrag{b}{\hskip -2.5cm $R_{-}(s)/ 10^{-2}$}
\begin{center}
\includegraphics[width=16cm,height=12cm]{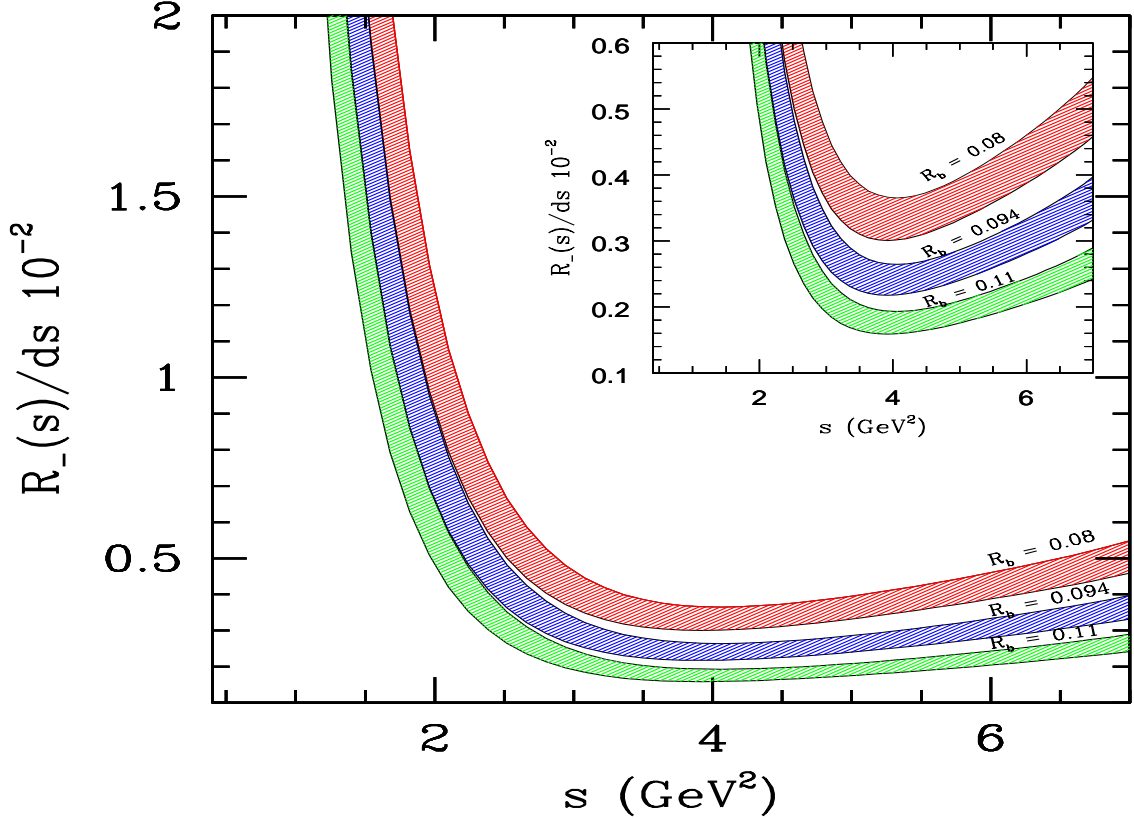}
\caption{The Ratio $R_{-}(s)$ with three indicated values of the CKM
ratio $R_{b} \equiv |V_{ub}|/|V_{tb}V_{ts}^*|$. The
bands reflect the theoretical uncertainty from $\zeta_{SU(3)}=1.3 \pm
0.06$  and $\xi^{(K^*)}_{{\perp}}(0)=0.28\pm 0.04$.}
\label{R-Vub}
\end{center}
\end{figure}

\hspace*{\parindent}
The measurement of exclusive $B \rightarrow \rho  \ell \bar{\nu}$ decays
is one of the major goals of B physics. It provides a good tool for
the extraction of  $|V_{ub}|$, provided the form factors can be either
measured precisely or calculated from first principles, such as the
lattice-QCD framework.  To reduce the non-perturbative uncertainty in the
extraction of $V_{ub}$, we propose to study the ratios of the
differential decay rates in $B \to \rho \ell \nu_\ell$ and $B \to K^* \ell^+
\ell^-$ involving definite helicity states.  These $s$-dependent ratios 
$R_{i}(s)$, $(i=0,-1,+1)$ are defined as follows: 
\begin{eqnarray}
R_{i}(s) ={{d\Gamma_{H_i}^{B \rightarrow  K^{*} \ l^{+}  l^{-}}/ds}
\over{d\Gamma_{H_i}^{B \rightarrow  \rho \ l \bar{\nu}}/ds}}
\label{Ratio} 
\end{eqnarray}
The ratio $R_{-}(s)$ suggests itself as the most interesting one, as the
form factor dependence essentially cancels (in the SU(3)-symmetry limit). From
this, one can measure
the ratio $\vert V_{ts} \vert/\vert V_{ub}\vert$. In
Fig.~(\ref{R-Vub}), we plot $R_{-}(s)$ for three representative values of the
CKM ratio $R_b = \vert V_{ub}\vert/\vert V_{tb} V_{ts}^* \vert =
\vert V_{ub}\vert/\vert V_{cb}\vert =0.08$, $0.094$, and $0.11$.  We also
 show the ratio $R_{0}(s)$, where the
form factor dependence does not cancel. For the LEET form factors used here,
the compounded theoretical uncertainty is shown by the shaded regions.  
This figure suggests that high statistics experiments may be able to
determine the CKM-ratio from measuring $R_{-}(s)$ (and $R_{0}(s)$) at a competitive level
compared to the other methods {\it en vogue} in experimental studies.

\begin{figure}
\psfrag{s}{ $\mathbf{s\ (GeV^2)}$}
\psfrag{R0}{\hskip -2cm $\mathbf{R_{0}(s)\times 10^{-2}}$}
\psfrag{Rb11}{$\mathbf{R_b=0.11}$}
\psfrag{Rb94}{$\mathbf{R_b=0.094}$}
\psfrag{Rb8}{$\mathbf{R_b=0.08}$}
\begin{center}
\includegraphics[width=16cm,height=12cm]{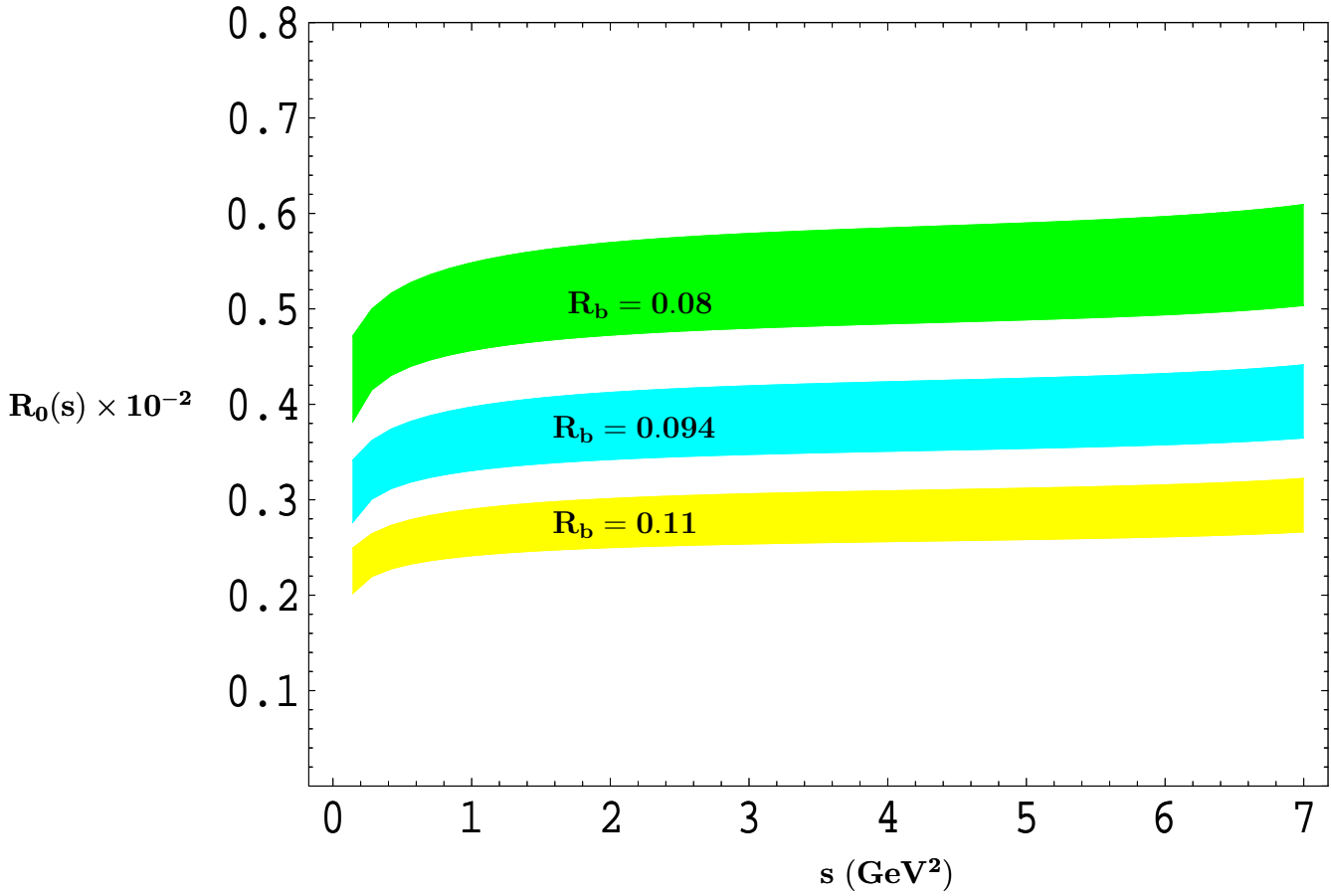}
\caption{The Ratio $R_{0}(s)$ with three indicated values of the CKM
ratio $R_{b} \equiv |V_{ub}|/|V_{tb}V_{ts}^*|$. The
bands reflect the theoretical uncertainty from $\zeta_{SU(3)}=1.3 \pm
0.06$  and $\xi^{(K^*)}_{{\perp}}(0)=0.28\pm 0.04$.}
\label{Rzero}
\end{center} 
\end{figure} 
\section{The Ratios $R_-(s)$ and $R_0(s)$ as Probes of
 New Physics in $B \rightarrow  K^{*} \ l^{+}  l^{-}$}
\hspace*{\parindent}
In order to look for new physics in $B \rightarrow  K^{*} \ell^{+}
\ell^{-}$,  we propose to study  the ratios $R_{0}(s)$ and $R_{-}(s)$,
introduced in the previous section. As well known, new physics can distort
the dilepton invariant mass spectrum and the forward-backward asymmetry 
in a non-trivial way. 

\begin{figure} 
\psfrag{a}{ $s\ (GeV^2)$}
\psfrag{b}{\hskip -2.5cm $R_{-}(s)/ 10^{-2}$}
\begin{center}
\includegraphics[width=12cm,height=9cm]{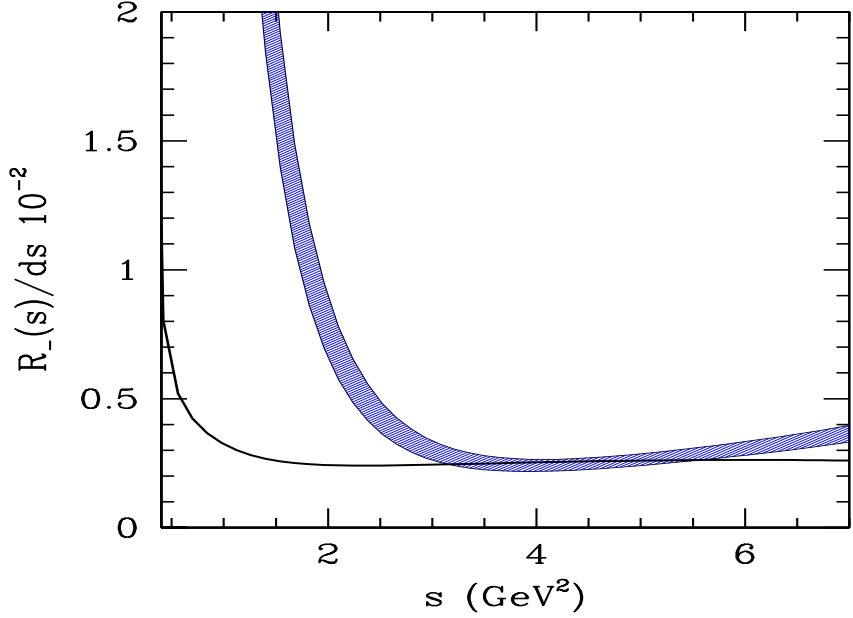}
\caption{The Ratio $R_{-}(s)$ with $|V_{ub}|/|V_{tb}V_{ts}^*|=0.094$ in the Standard Model and in SUGRA, with $(r_7,\ r_8)=( -1.2,\ -1)$, $\zeta_{SU(3)}=1.3$ and $\xi^{(K^*)}_{{\perp}}(0)=0.28$  represented, respectively, by the shaded 
area and the solid curve. The shaded area depicts the theoretical
uncertainty  $\zeta_{SU(3)}=1.3 \pm 0.06$  and
 $\xi^{(K^*)}_{{\perp}}(0)=0.28\pm 0.04$.}  
\label{R-SMetSUGRAneg}
\end{center}
\end{figure}

\begin{figure}
\psfrag{s}{ $\mathbf{s\ (GeV^2)}$}
\psfrag{R0}{\hskip -2cm $\mathbf{R_{0}(s)\times 10^{-2}}$}
\psfrag{Rb94}{$\mathbf{R_b=0.094}$} 
\begin{center}
\includegraphics[width=12cm,height=9cm]{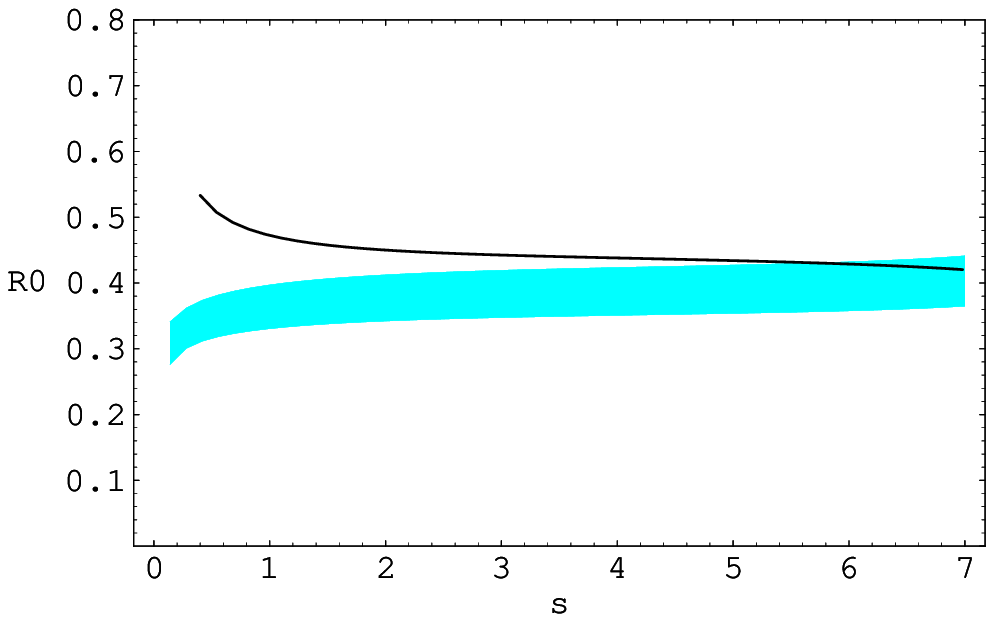}
\caption{The Ratio $R_{0}(s)/$ with
$|V_{ub}|/|V_{tb}V_{ts}^*|=0.094$ in the Standard Model and in 
SUGRA, with $(r_7,\ r_8)=( -1.2,\ -1)$, $\zeta_{SU(3)}=1.3$ and
  $\xi^{(K^*)}_{{\perp}}(0)=0.28$  represented, respectively, by the
shaded area and the solid curve. The shaded area depicts the theoretical
uncertainty  $\zeta_{SU(3)}=1.3 \pm 0.06$  and 
 $\xi^{(K^*)}_{{\perp}}(0)=0.28\pm 0.04$.}
\label{R0SM&errorneg}
\end{center}
\end{figure}

To illustrate generic SUSY effects in ${B \rightarrow  K^{*}
\ell^{+} \ell^{-}}$, 
we note that the Wilson coefficients $C^{\bf eff}_7$, $C^{\bf eff}_8$,
$C_9$ and $C_{10}$ receive additional contributions from the supersymmetric
particles. We incorporate these effects by assuming that the ratios of the
Wilson coefficients in these theories and the SM deviate from 1. These ratios
for $k= 7, 8, 9, 10$ are defined as follows:
\begin{eqnarray}
r_{k}(\mu) = {C_{k}^{SUSY}\over C_{k}^{SM}}~.
\label{rk} 
\end{eqnarray}
\noi They depend on the renormalization scale (except for $C_{10}$),
for which we take $\mu =m_{b,pole}$. For the sake of illustration,
we use representative values for the large-$\tan \beta$  SUGRA model,
in which the ratios $r_7$ and $r_8$ actually change their signs. The
supersymmetric effects on the other two Wilson coefficients $C_9$ and
$C_{10}$ are generally small in the SUGRA models, leaving $r_9$ and $r_{10}$
practically unchanged from their SM value. To be specific, we take 
\footnote{We thank Enrico Lunghi for providing us with these numbers.}
\begin{eqnarray}
r_{7} = -1.2,\ \ r_{8} = -1,\ \    \ r_{9} = 1.03,\ \ r_{10} = 1.0~.
\label{rnegativ}
\end{eqnarray} 
In Figs.~(\ref{R-SMetSUGRAneg}) and (\ref{R0SM&errorneg}), we present a
comparative study of the SM and SUGRA partial distribution 
for $H_{-}$ and $H_{0}$, respectively. In doing this, we also show the
attendant theoretical uncertainties for the SM, worked out in the LEET
approach in this paper. For
these distributions, we have used the form factors from 
\cite{Ali:1999mm} with the SU(3)-symmetry breaking parameter taken in the
range $\zeta_{SU(3)}=1.3\pm 0.06$.
Figs.~(\ref{R-SMetSUGRAneg}) and (\ref{R0SM&errorneg}) illustrate
clearly that despite non-perturbative uncertainties, it is possible, in
principle,  in the low $s$ region  to
distinguish between the SM and a SUGRA-type models, provided the ratios
$r_k$ differ sufficiently from 1.
%
\section{Summary and Concluding Remarks}
\hspace*{\parindent}
 Summarizing briefly our results,  we have reported an
$O(\alpha_s)$-improved analysis of the various helicity components
in the decays $B \to K^* \ell^+ \ell^-$ and $B \to \rho \ell \nu_\ell$,
carried out in the context of the Large-Energy-Effective-Theory.
The underlying symmetries in the large energy limit lead to an enormous
simplification as they reduce the number of independent form factors
in these decays. The LEET-symmetries are broken by QCD
corrections, and we have calculated the helicity components
implementing the $O(\alpha_s)$ corrections. The results presented here
make use of the form factors calculated in the QCD sum rule approach. The
LEET form factor $\xi_\perp^{K^*}(0)$ is constrained by current data on $B
\to K^* \gamma$. As the theoretical analysis is restricted to the lower
part of the dilepton invariant mass region in $B \to K^* \ell^+ \ell^-$,
typically $s < 8$ GeV$^2$, errors in this form factor are
not expected to severely limit theoretical precision. This implies that
distributions involving the $H_{-}(s)$ helicity component can be
calculated reliably. Precise measurements of the two LEET form factors
$\xi_{\perp}^{\rho}(s)$ and $\xi_{\parallel}^{\rho}(s)$ in
the decays $B \to \rho \ell \nu_\ell$ can be used to largely reduce the
residual model dependence. With the assumed form factors, we have worked
out a number of single and double (Dalitz) distributions in $B \to \rho
\ell \nu_\ell$, which need to be confronted with data. An analysis of the
decays $B \to K^* \ell^+ \ell^-$ is also carried out in the so-called
transversity basis. We have compared the LEET-based amplitudes in this
basis with the data currently available on $B \to K^* J/\psi(\to \ell^+
\ell^-)$ and find that the short-distance based transversity amplitudes
are very similar to their long-distance counterparts.  We also show the
$O(\alpha_s)$ effects on the forward-backward asymmetry, confirming
essentially the earlier work of Beneke, Feldmann and Seidel
\cite{Beneke:2001at}. Combining the analysis of the decay modes 
$B \to K^* \ell^+ \ell^-$ and $B \to \rho \ell \nu_\ell$, we show that
the ratios of differential decay rates involving definite helicity states, 
$R_{-}(s)$ and $R_{0}(s)$, can be used for testing the SM precisely.
We work out the dependence of these ratios on the CKM matrix elements
$\vert V_{ub}\vert/\vert V_{ts}\vert$.
 We have also analyzed possible effects on these
ratios from New Physics contributions, exemplified by representative
values for the effective Wilson coefficients in the large-$\tan \beta$ SUGRA
models. The
main thrust of this paper lies, however, on showing that the currently
prevailing theoretical uncertainties on the SM distributions in $B \to K^*
\ell^+ \ell^-$ can be largely reduced by using the LEET approach and data
on $B \to K^* \gamma$ and $B \to \rho \ell \nu_\ell$ decays. Finally, we
remark that the current experimental limits on $B \to (X_s,K^*) \ell^+
\ell^-$ decays (and the observed $B \to K \ell^+ \ell^-$ decay)
\cite{bellebsll,babarbsll,Affolder:1999eb,Anderson:2001nt}
are already probing the SM-sensitivity. With the integrated luminosities   
over the next couple of years at the $B$ factories, the
helicity analysis in  $B \to \rho \ell \nu_\ell$
and $B \to K^* \ell^+ \ell^-$ decays presented here can be
carried out experimentally.
 \pagestyle{plain}
\section*{Acknowledgements}
\hspace*{\parindent} A.~S.~S. would like to thank Thorsten Feldmann for
several helpful discussions, and the German Academic Exchange Service
(DAAD) and DESY for financial support. A.A. would like to thank Tony Sanda
and Mikihiko Nakao for illuminating discussions, and the latter
also for sharing the BELLE results and projections on rare $B$ decays.
We would like to thank Gustav Kramer and Amjad Gilani for 
pointing out some errors in  the earlier
version of this manuscript.

\end{document}